\documentclass[aps,prx,twocolumn,superscriptaddress]{revtex4-2}
\usepackage{amssymb,amsmath,amsfonts,graphicx,epsf,dcolumn,bm}
\usepackage{color}

\begin{document}

\title{Generalized entropy production in collisionless plasma flows and turbulence}

\author{Vladimir Zhdankin}
\email{vzhdankin@flatironinstitute.org}
\affiliation{Department of Astrophysical Sciences, Princeton University, 4 Ivy Lane, Princeton, NJ 08544, USA}
\affiliation{Center for Computational Astrophysics, Flatiron Institute, 162 Fifth Avenue, New York, NY 10010, USA}

\date{\today}

\begin{abstract}
Collisionless plasmas exhibit nonthermal and anisotropic particle distributions after being energized; as a consequence, they enter a {state of low Boltzmann-Gibbs (BG) entropy} relative to the thermal state. The Vlasov equations predict that in a collisionless plasma with closed boundaries, {BG} entropy is formally conserved, along with an infinite set of other Casimir invariants; this provides a seemingly strong constraint that may explain how plasmas maintain low entropy. Nevertheless, {it is commonly believed that entropy production is enabled by} phase mixing or nonlinear entropy cascades. The question of whether such anomalous entropy production occurs, and of how to characterize it quantitatively, is a fundamental problem in plasma physics. We construct a new theoretical framework for characterizing entropy production (in a generalized sense) based on a set of ideally conserved {dimensional quantities} derived from the Casimir invariants{; {these are referred to as the ``Casimir momenta'' and they} generalize the BG entropy}. The growth of the Casimir momenta relative to the average particle momentum indicates entropy production. We apply this framework to quantify entropy production in particle-in-cell simulations of laminar flows and turbulent flows driven in relativistic plasma, where efficient nonthermal particle acceleration is enabled. We demonstrate that a large amount of anomalous entropy is produced by turbulence despite nonthermal features. The Casimir momenta grow to cover a range of energies in the nonthermal tail of the distribution, and we correlate their growth with spatial structures. These results have implications for reduced modeling of nonthermal particle acceleration and for diagnosing irreversible dissipation in collisionless plasmas such as the solar wind and Earth's magnetosphere. {Dimensional representations of generalized entropy analogous to the Casimir momenta may be useful for other problems in statistical physics.}
\end{abstract}


\maketitle

\section{Introduction} \label{sec:intro}

Entropy is a fundamental quantity that underlies statistical physics, by characterizing the number of microscopic configurations that are consistent with the macroscopic properties of a system. The production of entropy is often interpreted as a signature of an irreversible process, since a system is unlikely to evolve from a ``more common'' to a ``less common'' microscopic configuration when acted on by statistically random forces. In the context of plasma physics, the kinetic entropy can be calculated from the particle momentum distribution function $f(\boldsymbol{x},\boldsymbol{p},t)$, with the standard form being the classical Boltzmann-Gibbs {(BG)} entropy $S(t) \equiv - \int d^3x d^3p f \log f$ {\citep[although many other types of entropies have also been devised; see, e.g., Refs.][]{renyi_1961,tsallis_1988, beck_cohen_2003}}. Entropy production is expected to accompany irreversible heating in a plasma. Shocks, magnetic reconnection, and turbulence are all examples of plasma processes that irreversibly convert energy from bulk motions or large-scale magnetic fields to internal plasma energy.

Collisionless plasmas provide a somewhat unusual setting in which the role and fate of entropy have not yet been established concretely, despite its fundamental nature. The difficulty lies in the fact that the absence of Coulomb collisions prohibits a direct route for the production of entropy. It is generally accepted that collisionless plasmas can be accurately described ``from first principles'' by the Vlasov-Maxwell equations, and that a variety of kinetic plasma mechanisms may cause irreversible energy dissipation, including Landau damping \citep{landau_1946}, Barnes damping \citep{barnes_1966}, ion-cyclotron resonance \citep[e.g.,][]{hollweg_isenberg_2002}, stochastic heating \citep[e.g.,][]{chandran_etal_2010}, etc. However, the Vlasov-Maxwell equations formally conserve entropy for suitable boundary conditions (e.g., a closed system), which seems to preclude the plasma from naturally attaining the maximum entropy state (i.e., a uniform thermal distribution). In fact, not only the BG entropy $S$, but also an infinite number of closely related integrals known as the Casimir invariants are conserved. {This plethora of conserved quantities suggests that the BG entropy is not a unique, or even the most appropriate, quantity for characterizing the entropy of a collisionless plasma. Conservation of the Casimir invariants would also seemingly impose} a strong constraint on the kinetic physics of the plasma (as will be described in Sec.~\ref{sec:background1}). Determining how irreversible dissipation occurs in a collisionless plasma subject to this constraint is a challenging mathematical problem, even for the simplest processes such as nonlinear Landau damping \citep[see, e.g.,][]{mouhot_villani_2011}. So, what happens to the overall entropy (and the Casimir invariants) as a collisionless plasma is energized?

Broadly speaking, one can conceive three resolutions to this question: (1) entropy may be produced by mechanisms beyond the Vlasov-Maxwell equations (i.e., collisions, however rare), (2) entropy may be effectively produced at macroscopic scales while being microscopically conserved, or (3) entropy may simply be conserved by the global distribution acquiring an appropriate {\it nonthermal} form. The lack of a clear answer to the question (applied to various energization processes) signifies a gap in our fundamental understanding of kinetic plasma physics. We now unpack these three {possible scenarios} in more detail.

{Scenario (1)} is that entropy conservation is broken by mechanisms ouside of the Vlasov-Maxwell equations, implying that the Vlasov-Maxwell equations are an incomplete description of a physical system. In particular, it has been suggested that the Vlasov equations must be supplemented by a collision operator to properly describe real systems, much like the Euler equations for an incompressible fluid must be supplemented by a viscous term to avoid the formation of singularities \citep[e.g.,][]{constantin_2007}. The collisions, no matter how insignificant, may then ultimately break entropy conservation. {For example, recent} phenomenological theories of kinetic turbulence suggest that entropy production may occur through an entropy cascade \citep[e.g.,][]{schekochihin_etal_2009, eyink_2018}, which forms fine-scale gradients in the particle distribution function and thus triggers collisional dissipation, even in the limit of an infinite mean free path. {The end result is thermal heating.} Recent numerical \citep[e.g.,][]{tatsuno_etal_2009, cerri_etal_2018, pezzi_etal_2018} and observational \citep[e.g.,][]{servidio_etal_2017} works indirectly support the existence of these types of phase-space cascades. {To verify this scenario directly, it is critical to have robust methods for characterizing violations of the Vlasov equation, which is a motivation of the present work.}

{Scenario (2)} is that entropy is conserved but effectively scrambled on microscopic scales, so that it appears to be systematically produced on macroscopic scales. In other words, there is apparent irreversible heating in the coarse-grained distribution of particles, but the fine-grained distribution retains memory of the initial entropy. In this context, the fine-grained distribution refers to a distribution that is binned on scales comparable to the characteristic kinetic scales of the plasma (skin depth, characteristic Larmor radius, or possibly the Debye length), while the coarse-grained distribution refers to one that is binned on larger scales (e.g., within the inertial range of turbulence, or comparable to the characteristic system size). Since in many applications, one is interested in the distribution function measured across macroscopic scales, entropy can be said to be effectively produced in this case. On this topic, we note that plasma echoes provide a plausible physical process by which a phase-space cascade of entropy may be inhibited, preventing collisions from playing any role \citep{gould_etal_1967, malmberg_etal_1968}. Stochastic plasma echoes have been shown to inhibit the momentum-space cascade in certain reduced plasma turbulence models \citep{schekochihin_etal_2016, parker_etal_2016, meyrand_etal_2019}. 

{Scenario (3)} is that entropy is conserved in a macroscopic sense. In this case, when energy is injected to the plasma, the global particle distribution must evolve systematically in a way that conserves entropy while admitting an increase in the average particle energy.  Thus, the distribution must depart from a state of maximum entropy and will become nonthermal. Intriguingly, nonthermal distributions are known to be ubiquitous in collisionless plasmas, as verified by countless laboratory experiments, space and astrophysical observations, and numerical simulations. It is then natural to ask, is there is a causal link between entropy conservation and the development of nonthermal particle distributions? Such ideas are worth considering based on recent particle-in-cell (PIC) simulations of magnetic reconnection that claim {BG} entropy is largely conserved on the whole, while being redistributed between spatial and kinetic degrees of freedom as a result of the plasma dynamics \citep{liang_etal_2019}.

{When comparing the three scenarios above, it is important to stress that in practice, it may be difficult to distinguish between Scenarios (1) and (2), because they can both yield similar manifestations on macroscopic scales (measurable by experiments or simulations). Violations of the Vlasov equation at coarse-grained scales could be attributed to either scenario, since the scrambling in Scenario (2) may be envisioned to occur on arbitrarily fine scales. In this situation, Scenarios (1) and (2) may be essentially indistinguishable at dynamical scales, and thus the question of which scenario dominates is mainly of {conceptual} nature. {In particular, we note that entropy non-conserving weak solutions of the Vlasov-Maxwell equations may be able to accurately describe the dynamics across a range of scales for either Scenario (1) or Scenario (2); we refer the interested reader to Ref.~\citep{bardos_etal_2020} for a discussion of the regularity conditions necessary for entropy conservation in weak solutions (for which global existence is known \citep{diperna_lions_1989}).} In contrast, the (full or partial) realization of Scenario (3) may have tangible implications, {as} will be described below.}

Determining which of the three preceding outcomes is realized (for various collisionless energization processes) is not only of fundamental interest, but has applications to modeling nonthermal particle acceleration and plasma anisotropies. If entropy is conserved, it acts as a valuable (and powerful) constraint on the form of the particle distribution that can be realized after energization. For example, this may give clues into why nonthermal particle distributions developed by relativistic magnetic reconnection and turbulence look so similar (in terms of power-law indices for energy distributions), both in 2D and in 3D domains \citep[e.g.,][]{sironi_spitkovsky_2014, comisso_sironi_2018, werner_uzdensky_2017, werner_uzdensky_2021}. On the other hand, if entropy is produced, then it may serve as a useful proxy for irreversible dissipation. In addition, characterizing entropy may yield insights on {the relevance of maximum-entropy states}, magnetic self-organization (a means of counteracting entropy production), and plasma phase transitions \citep{jara-almonte_ji_2021}. Despite these motivations, there are relatively few studies that attempt to directly measure kinetic entropy or related proxies in first-principles simulations of plasmas \citep[e.g.,][]{birn_etal_2006, dai_wong_2016, liang_etal_2019, liang_etal_2020, du_etal_2020, jara-almonte_ji_2021}. This is due not only to numerical issues (such as high resolution demands), but also conceptual issues such as how to meaningfully normalize entropy~\citep{liang_etal_2019}.

In this work, we construct a new theoretical framework for characterizing entropy production in collisionless plasmas, by utilizing the more general set of Casimir invariants. We propose that entropy production can be quantified robustly via an (infinite) set of characteristic {\it Casimir momenta} that represent the shape of the particle distribution in phase space. This framework opens up a novel approach to determining to what extent entropy is conserved at various energetic and spatial scales in a plasma. It also avoids normalization and uniqueness issues that are inherent to approaches based on the classical BG entropy. {Dimensional representations of generalized entropy analogous to the Casimir momenta may be useful for other problems in statistical physics beyond plasma physics.}

While the previous discussion is applicable to all collisionless plasma energization processes, as a proof of concept, we apply the framework to characterize anomalous entropy production in PIC simulations of two-dimensional (2D) relativistic plasma turbulence. Recent papers applied PIC simulations to demonstrate that relativistic turbulence naturally produces nonthermal particle distributions, with a power-law tail that extends across a broad range of energies \citep{zhdankin_etal_2017, comisso_sironi_2018}. The mechanism of nonthermal particle acceleration for high-energy particles resembles second-order Fermi acceleration \citep{comisso_sironi_2019, wong_etal_2020}. More rigorously, quasi-linear theory predicts diffusive acceleration from gyroresonant (or resonance-broadened) interactions between particles and various plasma modes at large scales \citep[e.g.,][and references therein]{demidem_etal_2020}. Magnetic reconnection inside of intermittent current sheets may also contribute to the energization of particles from the thermal population \citep{comisso_sironi_2018}. In addition to particle acceleration, the turbulent energization process produces both a systematic \citep{comisso_etal_2020} and a stochastic \citep{zhdankin_etal_2020, nattila_beloborodov_2021} energy-dependent anisotropy in the particle momentum distribution, which has implications for the radiative signatures of high-energy astrophysical systems. This rich nonthermal landscape makes relativistic turbulence an ideal testing ground for understanding the competition between entropy production and entropy conservation in a collisionless plasma.

Our numerical analysis demonstrates the violation of entropy conservation in PIC simulations of relativistic plasma turbulence. We rule out {Scenario (3)} outlined above for the fate of entropy, {revealing} that entropy is not conserved at macroscopic scales in a turbulent plasma, despite nonthermal features that may naively be expected to counteract entropy production. {In addition, we} place {strong} limits on {Scenario (2)} (fine-grained entropy conservation), {with Scenario (1) argued to be the most probable situation}. We then suggest that the Casimir momenta may be used as a diagnostic to gain insights into the irreversible dissipative processes operating in the plasma.

This paper is organized as follows. In Section~\ref{sec:background}, we describe the theoretical framework, by introducing the Casimir invariants and showing how a set of characteristic Casimir momenta can be extracted as a proxy for entropy production. To guide the reader, we present some basic mathematical properties of the Casimir momenta and highlight their qualitative similarities with {BG} entropy. In Section~\ref{sec:methods}, we describe the numerical implementation of the diagnostics for measuring the Casimir momenta and the PIC simulation setups. In Section~\ref{sec:laminar}, we describe results from PIC simulations of laminar flows in {three different configurations} as a benchmark for the diagnostics, showing the generalized entropy (i.e., Casimir momenta) to be conserved to a high degree relative to the injected energy. {Each of the laminar flows thus follow either Scenario (2) or (3) above.} In Section~\ref{sec:turb}, we describe results from the PIC simulation of 2D relativistic turbulence, showing that the Casimir momenta grow significantly due to irreversible energy dissipation. {Turbulence is thus best represented by Scenario (1), although we comment on the viability of Scenario (2).} Finally, in Section~\ref{sec:conclusions}, we summarize our conclusions, {discuss future outlook, and make connections to other problems in statistical physics}. 

\section{Theoretical Framework} \label{sec:background}

\subsection{Casimir invariants as a generalization of kinetic Boltzmann-Gibbs entropy} \label{sec:background1}

In this work, we {consider} the relativistic Vlasov-Maxwell equations{, which we assume is a reasonable model of a strictly collisionless plasma. The} Vlasov equation for a given particle species is
\begin{align}
\partial_t f + \boldsymbol{v}\cdot\nabla f + \boldsymbol{F} \cdot \frac{\partial f}{\partial \boldsymbol{p}} &= 0 \, , \label{eq:vlasov}
\end{align}
where $f(\boldsymbol{x},\boldsymbol{p},t)$ is the {fine-grained} particle distribution function (normalized such that $\int d^3p d^3x f = N$ is the total number of particles in the system),~$\boldsymbol{v} = \boldsymbol{p}c/ (m^2c^2+p^2)^{1/2}$ is the particle velocity (with $m$ the particle mass), and~$\boldsymbol{F}(\boldsymbol{x},\boldsymbol{p},t)$ is a {phase-space conserving} force field (satisfying $\partial/\partial\boldsymbol{p} \cdot \boldsymbol{F} = 0$). In the case of a plasma, $\boldsymbol{F}$ includes the Lorentz force and possible external forces{; however, the following theoretical framework is applicable to general forms of $\boldsymbol{F}$ and thus can be applied more broadly (e.g., to the gravitational Vlasov-Poisson equations)}. Eq.~\ref{eq:vlasov} can be applied to any particle species in the plasma; in the following theoretical discussion, we do not make any distinction between the various species (electrons, positrons, ions, etc.).

{Before proceeding, we note that while the relativistic Vlasov-Maxwell equations are widely applied in theoretical physics, there is not yet an established proof in the literature of the global existence and uniqueness of its solutions under generic conditions. Global existence and/or uniqueness were proven only under limiting assumptions, such as weak topologies \citep{diperna_lions_1989} or velocities bounded away from the speed of light \citep{glassey_strauss_1986}. This leaves open the possibility that Eq.~\ref{eq:vlasov} is incomplete. If this is so, then Eq.~\ref{eq:vlasov} may need to be supplemented by a collision operator, which prevents singularity formation and breaks entropy conservation at microscopic scales. Another foundational issue is that Eq.~\ref{eq:vlasov} assumes that the distribution $f$ can be defined on arbitrarily fine-grained scales (in the 6D phase space) and formally contains an infinite number of particles ($N \to \infty$), while real systems must be coarse-grained at some level and necessarily have a finite $N$. PIC simulations only provide approximate solutions to Eq.~\ref{eq:vlasov}; numerical discretization can then act both as a coarse-graining operator and as a collision operator, allowing violations of Eq.~\ref{eq:vlasov} to arise at small scales. These mathematical issues may limit the use of Eq.~\ref{eq:vlasov} for rigorously describing all aspects of the plasma, but we believe that Eq.~\ref{eq:vlasov} is sufficient for exploring the essential physical concepts described in the present paper. Indeed, the theoretical framework discussed in this paper provides a means for precisely characterizing departures of physical systems from the predicted Vlasov dynamics.}

A basic feature of Eq.~\ref{eq:vlasov} is that, under suitable boundary conditions, it conserves an infinite number of kinetic integrals known as the Casimir invariants \citep[see, e.g.,][]{morrison_1987, diperna_lions_1989, morrison_pfirsch_1990, ye_morrison_1992, rocha_etal_2005}. These Casimir invariants are obtained from the functional given by
\begin{align}
{\mathfrak C}_g(f) \equiv \frac{1}{N} \int d^3x d^3p g(f) \, , \label{eq:casg}
\end{align}
where $g(f)$ is any differentiable function of $f$, subject to the conditions described below. It is straightforward to demonstrate conservation of ${\mathfrak C}_g(f)$ by using Eq.~\ref{eq:vlasov} to compute the derivative explicitly:
\begin{align}
\frac{d{\mathfrak C}_g}{dt} &= \frac{1}{N} \int d^3x d^3p \frac{dg}{df} \partial_t f \nonumber \\
&= - \frac{1}{N} \int d^3x d^3p \frac{dg}{df} \left[ \boldsymbol{v}\cdot\nabla f + \boldsymbol{F} \cdot \frac{\partial f}{\partial \boldsymbol{p}} \right] \nonumber \\
&= -  \frac{1}{N}\int d^3x d^3p \left[ \boldsymbol{v}\cdot\nabla g + \boldsymbol{F} \cdot \frac{\partial g}{\partial \boldsymbol{p}} \right] \nonumber \\
&= - \frac{1}{N} \int d^3x d^3p \left[ \nabla \cdot \left( \boldsymbol{v} g \right) + \frac{\partial}{\partial \boldsymbol{p}} \cdot \left( \boldsymbol{F} g \right) \right] \nonumber \\
&= - \frac{1}{N} \int d^3p d\boldsymbol{S}_x \cdot \boldsymbol{v} g - \frac{1}{N} \int d^3x d\boldsymbol{S}_p \cdot \boldsymbol{F} g  = 0 \, .
\end{align}
Here, we used the divergence theorem to express the volume integrals as bounding surface integrals in position space (surface $\boldsymbol{S}_x$) and in momentum space (surface $\boldsymbol{S}_p$), and assumed boundary conditions that make them vanish. For example, if $f$ is periodic (or closed) in space, smooth at $p = 0$, and $\to 0$ as $p \to \infty$, then ${\mathfrak C}_g$ will be conserved as long as $g$ is an increasing function of $f$ and has a converging integral in momentum space. In addition, $\boldsymbol{F}$ must be non-singular as $p \to 0$ and not grow too strongly with $p$.

By choosing appropriate $g$, the Casimir invariants include the number of particles ($g = f$) and the BG entropy ($g = - f \log{f}$) as special cases. There are, however, an infinite number of other conserved quantities. Physically, conservation of the Casimir invariants is associated with incompressibility of the distribution in 6D phase space. Each level set of $f$ is constrained to have a fixed filling fraction; in other words, the number of occurrences for any given value of $f$ in the 6D phase space will stay constant in time. Intuitively, this arises from the fact that all particles in a given parcel of the distribution must evolve along the same phase space trajectory if the dynamics are deterministic (and thus reversible). As a consequence, the shape of $f$ in a given dimension can only be changed if the shape is also changed in another dimension to balance out the variation (for example, spatial structure of $f$ may be redistributed into momentum space structure, or vice-versa).

{It is important to note that while the Vlasov equation predicts ${\mathfrak C}_g(f)$ to be conserved when calculated from the fine-grained distribution $f$, there is no equivalent conservation law for ${\mathfrak C}_g(\tilde{f}_{\ell,\wp})$, where $\tilde{f}_{\ell, \wp}(\boldsymbol{x},\boldsymbol{p},t)$ is the coarse-grained distribution obtained by discretizing the distribution on a grid with bin sizes $\ell$ in position space and $\wp$ in momentum space \citep[alternatively, one can apply a smoothing operator with arbitrary kernel on $f$, as in, e.g., ][]{eyink_2018}. Thus, the ideal Vlasov dynamics allow conservation of the Casimir invariants to be broken at any coarse-grained scale, in principle, even if they are conserved at fine-grained scales. This is a difference from other conserved quantities such as particle number and total energy. The values of $\ell$ and $\wp$ needed to accurately represent the fine-grained distribution for a given Vlasov problem are {\it a priori} unknown.}

In this work, we treat the Casimir invariants as a generalization of {BG} entropy. If the Casimir invariants are conserved, then {the BG} entropy is necessarily conserved. Although one can envision special situations where {BG} entropy is conserved while the Casimir invariants are not conserved, there is no {{\it a priori}} reason to single out the {BG} entropy as a preferred conserved quantity. Instead, all the Casimir invariants must be treated on equal footing. In the remainder of the paper, we will use the term ``entropy'' to refer to this notion of ``generalized entropy'', while referring to the classical entropy $S$ as the {BG} entropy.

\subsection{General comments about Casimir invariants}

To build the reader's intuition, we make a few general comments about the Casimir invariants. First, we note that for an unperturbed ($\boldsymbol{F} = 0$) collisionless plasma, any uniform particle momentum distribution constitutes an equilibrium. Thus, although the isotropic thermal (Maxwell-J\"{u}ttner) distribution is a ``maximum entropy state'' in terms of the {BG} entropy, there is no {\it a priori} reason to expect such a state to be reached. This can be interpreted as a consequence of the vast degeneracy between the {BG} entropy and the other Casimir invariants.

Density fluctuations, temperature fluctuations, nonthermal particle populations, and momentum anisotropies all contribute to the Casimir invariants (as will be further described in Sec.~\ref{sec:casspatial}). In terms of {BG} entropy, these effects would decrease entropy relative to the reference thermal state (i.e., a Maxwell-J\"{u}ttner distribution with the same mean energy). 

In contrast, any operations that preserve the local symmetry of the distribution in 6D phase space do not contribute to the Casimir invariants, since these do not affect the overall shape of the distribution. For example, local shifts of the distribution in physical space or momentum space (corresponding to advection or bulk flows) do not contribute.  Likewise, local rotations of an anisotropic distribution do not contribute.

If one demands the distribution to always be isotropic and uniform, $f(\boldsymbol{x},\boldsymbol{p},t) = f(p,t)$, then there are insufficient degrees of freedom to evolve the distribution while conserving all of the Casimir invariants. Such an evolution would need to be described by a 1D force, $\boldsymbol{F}(\boldsymbol{p},\boldsymbol{x},t) = F(p,t) \hat{\boldsymbol{p}}$, which can satisfy the $\partial/\partial\boldsymbol{p} \cdot \boldsymbol{F} = 0$ condition while simultaneously remaining non-singular as $p \to 0$ if and only if $F = 0$. Thus, a 2D phase space (either one spatial dimension and one momentum dimension, or two momentum dimensions) is a minimal requirement for self-consistent evolution. A simple example of a uniform distribution (2D in momentum space) that conserves all of the Casimir invariants by exploiting anisotropy is given in the Appendix~\ref{sec:appendix}.

Finally, we make an important comment about the physical dimensions of the Casimir invariants. Since $f$ is a dimensional quantity (having units of {inverse phase volume --- i.e.,} inverse length cubed and inverse momentum cubed), the Casimir invariants will also be dimensional, with the dimensions determined by the corresponding function $g$. This presents an issue in the context of the kinetic {BG} entropy $S$, which does not have well-defined dimensions{, in the sense that} it involves a logarithm of $f$ {and thus the value is not consistent under a change of physical units (or alternatively, the normalization factor of $f$). The BG entropy is therefore} nontrivial to normalize in a meaningful way{{, and has} no natural zero point ({leading to the uncomfortable situation} that varying $N$ will preserve the Vlasov equation, but cause $S$ to shift).} {While one can arrive at physical conclusions by considering changes in $S$ rather than the absolute value of $S$}, this limits the usefulness of the {BG} entropy for {interpreting the level of entropy production in} a plasma \citep[see, e.g., discussion in][]{liang_etal_2019}. {These considerations motivate contemplating} other choices of ${\mathfrak C}_{g}$ that do have well-defined dimensions, as an alternative quantity. This leads us to introduce the Casimir momenta {as a generalized, dimensional characterization of entropy} in the next subsection.

\subsection{Characteristic Casimir momenta} \label{sec:pchar}

While the complete set of Casimir invariants provides the theoretical foundation for this work, it is necessary in practice to narrow the scope by selecting the Casimir invariants with the most practical utility. In the remainder of the paper, we will focus on Casimir invariants associated with power-law functions, $g = f^\chi$, where $\chi > 0$ is an arbitrary index, and denote the corresponding functional by
\begin{align}
{\mathcal C}_{\chi}(f) \equiv \frac{1}{N} \int d^3x d^3p f^\chi \, , \label{eq:cas}
\end{align}
where the subscript is now a real number rather than a function. {The Casimir invariants given by Eq.~\ref{eq:cas} are found in the definitions of Renyi entropy \citep{renyi_1961} and Tsallis entropy \citep{tsallis_1988}.} Since any choice of smooth function for $g$ can be expanded in a Taylor series, ${\mathcal C}_i$ for integer $i \in \mathbb{N}$ forms a basis for the other Casimir invariants. Intuitively, large values of $\chi$ are more sensitive to the bulk (low-energy part) of the distribution, while small values of $\chi$ are more sensitive to high energies; ${\mathcal C}_{\chi}$ will diverge at sufficiently small (or negative) values of $\chi$.

One can extract a quantity that has units of momentum times length (i.e., angular momentum) by raising ${\mathcal C}_{\chi}$ to the appropriate power. Specifically, the quantity ${\mathcal C}_{\chi}^{-1/3(\chi-1)}$ has these units. For a uniform thermal distribution, this expression evaluates to $\sim n_0^{-1/3} \langle p \rangle$ (as will be shown in Sec.~\ref{sec:casiso}), where $n_0$ is the average particle number density and $\langle p \rangle$ is the average (thermal) momentum. In the applications that motivate this work, $n_0$ is fixed in time by particle number conservation, while $\langle p \rangle$ may increase due to energy injection into the system. {Thus, to acquire a dimensional representation of entropy that can be directly compared to injected energy,} we factor out the length dimension and define the characteristic {\it Casimir momenta}:
\begin{align}
p_{{c},\chi}(f) &\equiv n_0^{1/3} {\mathcal C}_{\chi}^{-1/3(\chi-1)}(f) \nonumber \\
&= n_0^{1/3} \left( \frac{1}{N} \int d^3x d^3p f^\chi  \right)^{-1/3(\chi-1)} \, . \label{eq:pchar}
\end{align}
Thus, from the set of Casimir invariants, we have derived the (ideally conserved) Casimir momenta $p_{c,\chi}$ as a dimensional quantity that can be compared to the average momentum $\langle p \rangle$ at any time.

The set of Casimir momenta $p_{{c},\chi}$ for varying $\chi$ is the primary quantity that will be used in this work as a ``generalized'' measure of entropy. The $p_{{c},\chi}$ give a detailed characterization of how well the Casimir invariants (and hence, entropy) are conserved, relative to any change in the average particle momentum $\langle p \rangle$. If the plasma heats up in a way such that the plasma is perfectly thermalized (i.e., the local distribution becomes thermal with uniform temperature and density), then $p_{{c},\chi}$ will increase in proportion to the instantaneous average momentum, $p_{{c},\chi} \propto \langle p \rangle$ for all $\chi$. If, on the other hand, the Casimir invariants are perfectly conserved, then $p_{{c},\chi}$ will remain constant in time for all $\chi$, requiring the development of a highly nonthermal distribution. In general, the dynamics will lie somewhere between these two extremes, with $p_{{c},\chi}$ departing from $\langle p \rangle$ in a way that may have a complicated dependence on $\chi$.

{The mathematical form of the Casimir momenta is similar to the ``exponential entropies'' of Ref.~\cite{campbell_1966}, further studied by Refs.~\cite{koski_persson_1992, tabass_etal_2016}. The exponential entropies were proposed as a measure of the extent of a 1D distribution. The Casimir momenta can be viewed as a further generalization for the higher-dimensional, mixed phase space considered by the Vlasov equation, where the exponential entropies would acquire physical dimension of phase space volume and thus are not directly relatable to the system energetics. One may also consider the Casimir momenta to be fundamentally equivalent to the Renyi and Tsallis entropies, since the underlying phase-space integral is identical. The advantage of the Casimir momenta over these previous representations of generalized entropy is that it takes a form that is physically interpretable for the Vlasov system.}

In the following {four subsections (Sections~\ref{sec:bg},\ref{sec:casiso}, \ref{sec:casspatial}, and \ref{sec:casglobal})}, we describe some basic analytical properties of the Casimir momenta. In particular, we show that the set of $p_{{c},\chi}$ share several of the same properties as the {BG} entropy $S$. The reader more interested in the numerical results may skip ahead to Sec.~\ref{sec:methods}.

\subsection{Connection to Boltzmann-Gibbs entropy} \label{sec:bg}

{While the index $\chi$ of the Casimir momentum can be chosen freely as a weight toward different regions of phase space (as will be described in Sec.~\ref{sec:casiso}), the case of $\chi = 1$ deserves special consideration. At this point, the Casimir invariant (Eq.~\ref{eq:cas}) represents particle number conservation and is unity, ${\mathcal C}_{1} = 1$, but the corresponding Casimir momentum $p_{{c},1}$ is formally undefined due to the exponent $-1/3(\chi-1)$ becoming singular. The Casimir momenta in the limit of $\chi \to 1$ evaluate to
\begin{align}
\frac{p_{{c},\chi\to 1}}{n_0^{1/3}}  &= \lim_{\chi\to 1} \left( \frac{1}{N} \int d^3x d^3p f^\chi \right)^{-1/3(\chi-1)} \nonumber \\
&= {f_{\rm ref}^{-1/3}} \lim_{\epsilon\to 0}  \left( \frac{1}{N} \int d^3x d^3p f e^{\epsilon \log{(f{/f_{\rm ref}})}} \right)^{-1/3\epsilon} \nonumber \\
&=  {f_{\rm ref}^{-1/3}}  \lim_{\epsilon\to 0}  \left[ \frac{1}{N} \int d^3x d^3p f \left(1 + \epsilon \log{\frac{f}{{f_{\rm ref}}}}\right) \right]^{-1/3\epsilon} \nonumber \\
&=  {f_{\rm ref}^{-1/3}  \lim_{\epsilon\to 0}  \left( 1 + \frac{\epsilon}{N} \int d^3x d^3p f \log{\frac{f}{f_{\rm ref}}} \right)^{-1/3\epsilon}} \nonumber \\
&=  {f_{\rm ref}^{-1/3}  \exp{\left(- \frac{1}{3 N} \int d^3x d^3p f \log{\frac{f}{f_{\rm ref}}} \right)}} \nonumber \\
&=  {\exp{\left(- \frac{1}{3 N} \int d^3x d^3p f \log{f} \right)}} \nonumber \\
&= e^{ S/3N} \, ,
\end{align} 
{where in the intermediate steps we introduced $f_{\rm ref}$, an arbitrary auxiliary constant with the same physical dimensions as $f$, to make the argument of the logarithm dimensionless.} Thus, {\it in the limit of $\chi \to 1$, the Casimir momenta reduce to the {BG} entropy $S = - \int d^3x d^3p f \log{f}$, acted on by an exponential operator that {restores} physical dimensions}. {In other words, the exponentiation negates the effect of the logarithm acting on $f$, which is responsible for {the usual issues involving the arbitrary normalization and zero point of} $S$.} Notably, this relationship reveals a procedure for extracting a dimensionally meaningful quantity from $S$.}

\subsection{Casimir momenta for uniform isotropic distributions} \label{sec:casiso}

Physically, the Casimir momenta $p_{{c},\chi}$ describe the typical width of the momentum distribution at a given value of $f$. In this sense, the index $\chi$ acts as a weight to capture different parts of the distribution (or different regions of the momentum space). In fact, there is a one-to-one mapping between $p_{c,\chi}$ and different values of $f$ for the idealized case of a uniform isotropic distribution that monotonically decreases with $p$. In this subsection, we illustrate the connection between $p_{{c},\chi}$ and different regions of momentum space by evaluating $p_{c,\chi}$ for the examples of a uniform thermal distribution and a nonthermal power-law distribution. 

The first example is the uniform Maxwell-J\"{u}ttner distribution, also known as the relativistic thermal distribution:
\begin{align}
f = \frac{n_0}{4\pi m^2 c T K_2 (m c^2/T)} \exp{\left( - \frac{\sqrt{m^2 c^4 + p^2 c^2}}{T} \right)} \, , \label{eq:mj}
\end{align}
where $K_2$ is the modified Bessel function of the second kind and $T$ is the temperature. For simplicity, we calculate $p_{c,\chi}$ separately in the ultra-relativistic and non-relativistic limits. In the ultra-relativistic limit ($T \gg m c^2$, relevant to the PIC simulations described later in this paper), Eq.~\ref{eq:mj} becomes
\begin{align}
f &\sim \frac{n_0 c^3}{8\pi T^3}  e^{-pc/T} \, .
\end{align} 
In this case, the Casimir momenta evaluate to
\begin{align}
p_{{c},\chi} &= n_0^{1/3} \left[ \frac{1}{N} \int d^3x d^3p  \left(  \frac{n_0 c^3}{8\pi T^3}  e^{-p c/T}\right)^{\chi}  \right]^{-1/3(\chi-1)} \nonumber \\
&= n_0^{1/3} \left[ \frac{4\pi}{n_0} \left(  \frac{n_0 c^3}{8\pi T^3} \right)^{\chi}  \int dp p^2  e^{-\chi p c/T} \right]^{-1/3(\chi-1)} \nonumber \\
&=  (8\pi)^{1/3} \chi^{1/(\chi-1)} T/c \nonumber \\
&= {\frac{(8\pi)^{1/3}}{3} \chi^{1/(\chi-1)} \langle p \rangle} \, \label{eq:pcasmj}
\end{align}
{where we used the fact that the average momentum $\langle p \rangle = 3 T/c$ in this ultra-relativistic limit}. Note that $\chi^{1/(\chi-1)}$ is a monotonically decreasing function, approaching $\infty$ as $\chi \to 0$ and $1$ as $\chi \to \infty$; it equals $e$ at $\chi = 1$. Thus, there is a one-to-one mapping between the set of $p_{{c},\chi}$ and different parts of the distribution $f$. In particular, $\chi \lesssim 1$ corresponds to momenta far in the tail of the distribution, while $\chi \gg 1$ corresponds to momenta close to the average momentum.

In the non-relativistic limit ($T \ll m c^2$), Eq.~\ref{eq:mj} becomes
\begin{align}
f &\sim {\frac{n_0}{(2\pi m T)^{3/2}}}  e^{-p^2/2mT} \, 
\end{align}
and the characteristic momenta can be shown to be
\begin{align}
p_{{c},\chi} &=  {(2\pi m T)^{1/2}} \chi^{1/2(\chi-1)} \nonumber \\
&= {\frac{\pi}{2}  \chi^{1/2(\chi-1)} \langle p \rangle } \,
\end{align}
{where we used $\langle p \rangle = (8 m T / \pi)^{1/2}$ in this limit.} Once again, $\chi^{1/2(\chi-1)}$ is a monotonically decreasing function, approaching $\infty$ as $\chi \to 0$ and $1$ as $\chi \to \infty$; the interpretation of $p_{c,\chi}$ is thus qualitatively similar to that for the ultra-relativistic case.

The next example is a uniform power-law distribution in momentum space, as would arise from nonthermal particle acceleration.  For the purposes of illustration, we assume sharp cutoffs of the distribution at some minimum momentum $p_{\rm min}$ and maximum momentum $p_{\rm max}$, with a large energetic range $p_{\rm max} \gg p_{\rm min}$, and a power-law index $a > 3$ in 3D momentum space so that the majority of particles reside near $p_{\rm min}$. In this case,
\begin{align}
f &\sim A_0 p^{-a} \, ,
\end{align}
where $A_0 = (a - 3) n_0/4 \pi p_{\rm min}^{3-a}$. The Casimir functionals evaluate to ${\mathcal C}_{\chi} = A_0^\chi (p_{\rm max}^{3-a \chi} - p_{\rm min}^{3-a \chi})/(3-a\chi)$. Thus, the main contribution to ${\mathcal C}_{\chi}$ is from $p_{\rm min}$ if $\chi > 3/a$ and from $p_{\rm max}$ if $\chi < 3/a$. As a consequence, $p_{{c},\chi} \sim p_{\rm min}$ for $\chi > 3/a$ and $p_{{c},\chi} \sim p_{\rm max}$ for $\chi < 3/a$, meaning that large $\chi$ capture momenta near the lower cutoff while small $\chi$ capture momenta near the upper cutoff\footnote{Strictly speaking, there is also a power of $p_{\rm min}$ from the normalization factor $A_0$, so the Casimir momenta at small $\chi$ evaluate to a hybrid scale, $p_{c,\chi} \sim p_{\rm min}^{(3-a) \chi/3(\chi-1)} p_{\rm max}^{(a\chi-3)/3(\chi-1)}$. But for sufficiently small $\chi$, the numerical prefactors cause $p_{c,\chi} \to p_{\rm max}$.}, with the critical index $\chi_{\rm crit} = 3/a$ being sensitive to both bounds. For typical situations of interest, relativistic particle acceleration leads to power-law indices in the energy distribution between $-1$ and $-4$, corresponding to $3 < a < 6$; this corresponds to critical indices covering the narrow range $1 > \chi_{\rm crit} > 1/2$. In our numerical analysis, we will choose $\chi$ to encompass this range.

In summary, $p_{{c},\chi}$ characterize the shape of the distribution, with a one-to-one mapping (at $p \gtrsim \langle p \rangle$) in the case of a uniform, isotropic distribution that declines monotonically with $p$. As demonstrated for a thermal distribution and a power law distribution, $\chi \lesssim 1$ corresponds to momenta that are far in the tail, while $\chi \gg 1$ corresponds to momenta that are close to the thermal momentum (or lower cutoff of the power law). The non-conservation of $p_{{c},\chi}$ at any values of $\chi$ may then be interpreted as the violation of entropy conservation in the corresponding region of phase space.

\subsection{Role of inhomogeneity and anisotropy in decreasing Casimir momenta} \label{sec:casspatial}

The presence of spatial inhomogeneity or momentum anisotropy acts to decrease entropy in a system. In much the same way, these features will decrease the Casimir momenta $p_{c,\chi}$ relative to the uniform, isotropic case. This is an important property of $p_{c,\chi}$ that we will demonstrate in this subsection. These effects are particularly relevant for the problem of turbulence since inhomogeneities arise from temperature or density fluctuations, and collisionless plasmas generally acquire (and maintain) an anisotropic distribution with respect to the magnetic field due to the anisotropy of the energization mechanisms.

{We first consider the role of spatial inhomogeneities. Consider a distribution decomposed into a uniform and a fluctuating part, $f(\boldsymbol{x},\boldsymbol{p}) = \overline{f}(\boldsymbol{p}) + \delta f(\boldsymbol{x},\boldsymbol{p})$, where $\overline{f} = \int d^3x f/V$ is the global distribution and $V$ is the integration volume. H\"{o}lder's inequality implies that $(\int d^3x f /V )^\chi \le \int d^3x f^\chi / V$ for {$\chi > 1$} and $(\int d^3x f /V )^\chi \ge \int d^3x f^\chi / V$ for {$0<\chi < 1$}. Integrating over momenta, this implies that the Casimir invariants computed from the global distribution obey ${\mathcal C}_{\chi}(\overline{f}) \le {\mathcal C}_{\chi}(f)$ for {$\chi > 1$} and ${\mathcal C}_{\chi}(\overline{f}) \ge {\mathcal C}_{\chi}(f)$ for {$0<\chi < 1$}. In both cases, this causes $p_{{c},\chi}$ to decrease from the value calculated with the uniform distribution $\overline{f}$ alone. As a result, {\it any nontrivial spatial structure will act to either decrease the Casimir momenta from the uniform case or, at best, keep it unchanged (as in the case of bulk flows or local rotations of the distribution)}.}

{To determine the effect of momentum anisotropy, next consider a uniform distribution decomposed into an isotropic part and an anisotropic fluctuation, $f(\boldsymbol{x},\boldsymbol{p}) = f_{\rm iso}(p) + \delta f(p, \theta, \phi)$, where $f_{\rm iso}(p) = \int d\phi d\cos{\theta} f / 4\pi$ and $\boldsymbol{p} = p(\cos{\phi}\sin{\theta} \hat{\boldsymbol{x}} + \sin{\phi}\sin{\theta}\hat{\boldsymbol{y}} + \cos{\theta} \hat{\boldsymbol{z}})$ defines the spherical coordinate system. For simplicity, we assume that $f$ is in the center-of-mass frame (otherwise a coordinate transform is needed).  H\"{o}lder's inequality implies that $(\int d\phi d\cos{\theta} f /4\pi )^\chi \le \int d\phi d\cos{\theta} f^\chi / 4\pi$ for {$\chi > 1$} and $(\int d\phi d\cos{\theta} f /4\pi)^\chi \ge \int d\phi d\cos{\theta} f^\chi / 4\pi$ for {$0<\chi < 1$}. Thus the Casimir invariants computed from the isotropic distribution obey ${\mathcal C}_{\chi}(f_{\rm iso}) \le {\mathcal C}_{\chi}(f)$ for {$\chi > 1$} and ${\mathcal C}_{\chi}(f_{\rm iso}) \ge {\mathcal C}_{\chi}(f)$ for {$0<\chi < 1$}. In both cases, this causes $p_{{c},\chi}$ to decrease from the value calculated with the isotropic distribution $f_{\rm iso}$ alone. Thus, {\it anisotropy decreases the Casimir momenta relative to the isotropic case}.}

Thus, we have shown that spatial inhomogeneities and momentum anisotropy both contribute to decreasing the Casimir momenta. This qualitative feature is analogous to the well-known effect of nontrivial (ordered) structure decreasing the {BG} entropy.

\subsection{Growth of global Casimir momenta with energy injection} \label{sec:casglobal}

The second law of thermodynamics dictates that systems will tend to evolve from low entropy to high entropy states (e.g., after energy is injected). Similarly, the Casimir momenta $p_{c,\chi}$ computed from the coarse-grained distribution have a tendency to grow after energy is injected into the system. In this subsection, we illustrate this by considering the evolution of $p_{c,\chi}$ computed from the global (spatially averaged) distribution $\overline{f}$, rather than the local distribution $f$. We will outline an argument that connects the growth of these global Casimir momenta with the overall heating.

Consider the global distribution $\overline{f}(\boldsymbol{p},t) \equiv \int d^3x f(\boldsymbol{x},\boldsymbol{p},t)/V$. Using Eq.~\ref{eq:vlasov}, the evolution of the global distribution is given by
\begin{align}
\partial_t \overline{f} &= - \frac{\partial}{\partial\boldsymbol{p}} \cdot \boldsymbol{\mathcal F} \, , \label{eq:vlasovcoarse}
\end{align}
where we defined $\boldsymbol{\mathcal F}(\boldsymbol{p},t) \equiv \int d^3x \boldsymbol{F} f/V$ {as the momentum-space force density}. The evolution of the Casimir invariants associated with $\overline{f}$ can then be expressed as
\begin{align}
\frac{d{\mathfrak C}_g(\overline{f})}{dt} &= \int d^3p g'(\overline{f}) \partial_t \overline{f} \nonumber \\
&= -  \int d^3p g'(\overline{f}) \frac{\partial}{\partial\boldsymbol{p}} \cdot \boldsymbol{\mathcal F} \nonumber \\
&= \int d^3p g''(\overline{f}) \frac{\partial \overline{f}}{\partial\boldsymbol{p}} \cdot \boldsymbol{\mathcal F} \, , \label{eq:cglobal}
\end{align}
where we used primes to indicate derivatives. For comparison, the overall ``heating'' rate $Q$ --- defined as the rate of increase of overall plasma energy --- has a similar form:
\begin{align}
\frac{Q}{V} &= \frac{d}{dt} \int d^3p d^3x \frac{E f}{V} = \int d^3p E \partial_t \overline{f} = \int d^3p \boldsymbol{v} \cdot \boldsymbol{\mathcal F} \, ,  \label{eq:qglobal}
\end{align}
where we defined the relativistic energy $E = (m^2 c^4 + p^2 c^2)^{1/2}$ and used $\partial E/\partial \boldsymbol{p} = \boldsymbol{v}$.

By comparing Eq.~\ref{eq:cglobal} with Eq.~\ref{eq:qglobal}, one can see that $\boldsymbol{\mathcal F}$ appears in the momentum-space integrals for both $d{\mathfrak C}_g(\bar{f})/dt$ and $Q$, but with different weights. As a result, one {may} expect the Casimir invariants for the {global} distribution to evolve in a similar way as $Q$. If energy is injected into the plasma, $Q > 0$, then {the product $v\boldsymbol{\mathcal F}$} must have a net positive $\hat{\boldsymbol{p}}$ component {via Eq.~\ref{eq:qglobal}}. In most realistic cases, $\overline{f}$ decreases monotonically with $p$, so $\partial \overline{f}/\partial p < 0$. {Although not rigorous, these considerations suggest that the term $\partial \overline{f} / \partial \boldsymbol{p} \cdot \boldsymbol{\mathcal F}$ will tend to be negative. Then,} as the plasma heats up, Eq.~\ref{eq:cglobal} implies that ${\mathfrak C}_g(\overline{f})$ will grow if $g'' < 0$, decline if $g'' > 0$, and stay constant if $g(f) = f$ (as it must to enforce particle number conservation). Interestingly, in terms of the Casimir momenta, this corresponds to $p_{{c},\chi}$ increasing after energy injection, for all $\chi > 0$. As a result, {\it energy injection will typically cause an increase of the Casimir momenta when measured from the global distribution (regardless of whether the Casimir invariants are locally conserved).}

This tendency for $p_{c,\chi}$ computed from $\overline{f}$ to increase with energy injection can be understood intuitively by the fact that spatial inhomogeneities (which will develop inevitably from realistic energy injection mechanisms) lead to a local reduction of entropy. To conserve the local entropy, this reduction must be compensated by an increase of entropy via the thermal structure of the distribution. When averaging the distribution over space, only the latter effect (changes in the thermal structure) is retained; thus a net increase in entropy (and the Casimir momenta) is expected. The numerical results in Sections~\ref{sec:laminar} and \ref{sec:turb} will confirm this expectation, by showing that $p_{c,\chi}$ computed from both the global distribution and the local (coarse-grained) distribution increase in time.

\section{Numerical methodology} \label{sec:methods}

We now proceed to apply the analytical framework described in Sec.~\ref{sec:background} to PIC simulations of laminar flows and turbulent flows in relativistic pair plasma. This is a means to demonstrate that the Casimir momenta $p_{{c},\chi}$ can be applied to real data to obtain conclusions about the amount of entropy produced in various regions of phase space, thus providing a proof of principle for the applicability of the methods to future problems. In this section, we first describe the implementation of the diagnostics for measuring $p_{c,\chi}$ in PIC simulations, and then describe the numerical setup of the PIC simulations.

\subsection{Measurement of the Casimir invariants}

In each PIC simulation, we construct a coarse-grained particle distribution for each species, from which we compute the Casimir invariants ${\mathcal C}_{\chi}$ via Eq.~\ref{eq:cas} for a representative set of indices, and then derive the Casimir momenta $p_{{c},\chi}$ via Eq.~\ref{eq:pchar}. In this work, we present results for $\chi \in \{ 1/3, 1/2, 3/4, 3/2, 2, 3 \}$. The Vlasov equations predict that the Casimir invariants are conserved for each particle species separately; we choose to show results acquired from the electron distribution, but confirmed that the positron distribution gives essentially identical results.

The coarse-grained particle distribution is binned on a Cartesian grid in both momentum space and physical space, with bin sizes $\Delta p_{i,\rm bin}$ in the three momentum directions ($i \in \{ x, y, z\}$) and $\Delta x_{\rm bin}$ in the two spatial directions\footnote{Alternatively, the distribution can be binned in spherical coordinates in momentum space. This may be advantageous for describing nonthermal distributions that span a broad range of energies, since bins can be chosen on a logarithmic energy scale. However, we find that the spherical coordinate singularities introduce additional numerical errors in measuring the Casimir invariants, so we avoid such a system in the present work.}. To measure the Casimir invariants accurately, it is essential to bin the distribution with sufficient resolution in momentum space, such that the peak of the distribution (near the average momentum $\langle p \rangle$) is resolved, while also having large enough bins to encompass nonthermal tails. In addition, there must be sufficient macroparticles per bin for the distribution to be smooth (otherwise the statistical noise manifests as artificial structure in the distribution, reducing $p_{c,\chi}$ from the physical value via the arguments in Sec.~\ref{sec:casspatial}). To achieve optimal resolution, we use an adaptive grid in momentum space by shifting and rescaling the coordinates with respect to the average momentum values in the spatial bin. Firstly, to remove broadening by large-scale flows, we shift the distribution toward the origin in momentum space by defining new momenta $\boldsymbol{p}' = \boldsymbol{p} - \langle \boldsymbol{p} \rangle_{\rm bin}$, where $\langle \boldsymbol{p} \rangle_{\rm bin}$ is the average particle momentum in the spatial bin. While not strictly necessary, this transformation to center-of-momentum coordinates improves the accuracy of the measurement in situations with high flow speeds. Secondly, we choose a momentum bin size (for each direction) that is a fixed fraction of the rms average momentum in the spatial bin at that time, $p_{i,{\rm rms, bin}} = (\langle  p_i^2\rangle_{\rm bin} - \langle p_i \rangle_{\rm bin}^2)^{1/2}$. This adaptive binning procedure allows the Casimir invariants to be measured accurately even with strong local variations in the plasma distribution and with significant heating of the plasma over time.

For the runs described in this paper, we resolve the distribution with either $32^2$ bins (laminar cases) or $64^2$ bins (turbulence case) in position space, and $256^3$ bins in momentum space. We choose a momentum bin size of $\Delta p_{i,\rm bin} = p_{i,{\rm rms, bin}}/4$, allowing the rms average momentum to be resolved by several cells. This leaves a factor of $p_{\rm max, bin}/p_{i,{\rm rms, bin}} = 32$ between the maximum momentum covered by the bins and the average momentum, sufficient to entirely cover the nonthermal tail of the distribution in the turbulence case (as will be described in Sec.~\ref{sec:turb}).

We performed a large number of simulations to test numerical convergence of the methods. This includes varying the number of particles per cell, the spatial resolution, and the binning resolution. The parameters chosen for the production simulations were informed by these numerical tests. We note that an insufficiently large value of $p_{\rm max, bin}$ causes ${\mathcal C}_{\chi}$ to artificially decrease, as portions of the distribution are missed, and thus $p_{{c},\chi}$ increases from the physical value if $\chi > 1$ and decreases if $\chi < 1$; we observed this effect in simulations with insufficiently large $p_{\rm max, bin}$ when testing the methods. We also confirmed that if $\Delta p_{i,{\rm bin}}$ is too small (or number of particles per cell too small) as to make the number of macroparticles per bin order unity, the statistical noise artificially decreases $p_{{c},\chi}$ from the converged value. {For reference, the dependence of the results on the spatial bin size $\Delta x_{\rm bin}$ is described in Appendix~\ref{sec:appendix2}.}

\subsection{Simulation setup}

We performed new simulations of driven laminar and turbulent flows using the PIC code {\em Zeltron} \citep{cerutti_etal_2013}. All simulations consider a relativistic electron-positron (pair) plasma, for numerical and theoretical simplicity. Although the dynamics will differ for a non-relativistic electron-ion plasma, we conjecture that the overall conclusions about entropy production may not be sensitive to the composition.

For simplicity, we focus on simulations that are 2D in physical space (neglecting the $z$ coordinate) and 3D in momentum space{; the higher momentum-space dimension is necessary to allow for out-of-plane electric currents and thus realistic dynamics.} The domain is thus a periodic box of area $L^2$. 2D setups allow the distribution to be coarse-grained onto a finer grid than is possible in 3D, which is essential for measuring Casimir invariants accurately at small and intermediate scales. We note that 2D and 3D setups yield similar nonthermal particle distributions in PIC simulations of both relativistic magnetic reconnection \citep[e.g.,][]{sironi_spitkovsky_2014, werner_uzdensky_2017, werner_uzdensky_2021} and relativistic turbulence \citep{comisso_sironi_2018, comisso_sironi_2019}, so we do not anticipate the basic conclusions in this work to be fundamentally different in 3D. 

We initialize particles from a uniform Maxwell-J\"{u}ttner distribution with an ultra-relativistic temperature, $\theta_0 \equiv T_0/m_e c^2 = 100$; we confirmed that the initial $p_{{c},\chi}$ measured in the simulations agree with the analytical values (Eq.~\ref{eq:pcasmj}) for the fiducial binning parameters and number of particles per cell (chosen to be $32$ per species). The simulations are permeated by a uniform mean magnetic field, $\boldsymbol{B}_0$, and the initial plasma beta based on this field is set to $\beta_0 \equiv 16\pi n_0 T_0/B_0^2 = 1/4$, where $n_0$ is the mean particle number density per species. This value of $\beta_0$ is sufficiently low to enable efficient nonthermal particle acceleration in the turbulence simulation. We note that the relativistic Alfv\'{e}n velocity is $v_A/c \equiv [\sigma/(1+\sigma)]^{1/2} \approx 0.63$, where $\sigma = 1/(2\beta)$ is the magnetization parameter in the relativistically hot limit. The simulation resolution is set so that the particle gyroradius $\rho_{e} = \langle p \rangle c/e B_0$ is initially resolved by 3 cells, $\Delta x_{\rm sim} = \rho_{e0}/3$. For reference, the initial value of the {relativistic} electron skin depth is $d_{e0} = 2 \rho_{e0}/(3\beta_0)^{1/2} \approx 2.3\rho_{e0}$ {and the Debye length is $\lambda_{D0}= (T_0/8\pi n_0 e^2)^{1/2} \approx d_{e0}/\sqrt{6} \approx 0.9 \rho_{e0}$}. The laminar flow simulations have $1024^2$ cells and thus $L/2\pi\rho_{e0} \approx 54$.  The turbulence simulation has $2048^2$ cells and $L/2\pi\rho_{e0} \approx 109$. Each coarse-graining bin contains $32^2$ cells in both setups.

The plasma is disturbed by an external force $\boldsymbol{F}_{\rm ext}$, which drives either a laminar flow (Sec.~\ref{sec:laminar}) or a turbulent flow (Sec.~\ref{sec:turb}). In this work, we focus on incompressible shear forces, leaving the case of compressive forces to future work \citep[for a discussion of the role of solenoidal versus compressive driving on turbulent particle acceleration, see][]{zhdankin_2021}. {For completeness, however, we also describe the case of collisionless damping of a density fluctuation in Appendix~\ref{sec:appendix3}, where it is shown analytically and numerically that the decay leads to a modest increase in $p_{c,\chi}$ via phase mixing.}

\section{Numerical results on laminar flows} \label{sec:laminar}

To benchmark the numerical diagnostics, we first apply them to a series of PIC simulations in which a laminar flow is driven at a single scale by a time-independent shear force:
\begin{align}
\boldsymbol{F}_{\rm shear}(x) = F_0 \sin{(k x)} \hat{\boldsymbol{y}} \, , \label{eq:fysin}
\end{align}
where $k = 2\pi/L$ is the wavenumber and we set amplitude of the force $F_0 = 0.6 k T_0$. We track the evolution of the system for times before any instabilities arise (either numerical or physical). Since there is negligible irreversible energy dissipation in these cases, entropy and the Casimir invariants are expected to be conserved. These cases thus serve to determine the adequate bin resolution, number of particles per bin, and so on required to verify conservation of $p_{c,\chi}$ to high accuracy.

\subsection{Neutral shear flow} \label{sec:neutral}

Before considering magnetized plasmas, it is useful to consider the dynamics of a collection of uncharged particles --- in other words, a collisionless relativistic gas driven by a shear force. This is described by the Vlasov equation with no Lorentz force:
\begin{align}
\partial_t f + v_x \partial_x f + F_{\rm shear}(x) \frac{\partial f}{\partial p_y} &= 0 \, , 
\end{align}
where $F_{\rm shear}(x)$ is the shear force given by Eq.~\ref{eq:fysin}; this can be easily simulated in PIC by setting the electric charge $q_s = 0$ \citep[see][for the analytical solution to this equation in the non-relativistic limit]{zhou_etal_2022}. The shear force drives bulk flows in the $\pm \hat{\boldsymbol{y}}$ directions, while streaming of the particles in $\pm \hat{\boldsymbol{x}}$ directions causes phase mixing. While not strictly irreversible, phase mixing {enables entropy production at coarse-grained scales} by scrambling the distribution in momentum space as particles sample the force to varying degrees while crossing the box at their transverse speed $v_x$. Particles return to their original state after crossing the box in $\hat{\boldsymbol{x}}$; while most particles do this on the light crossing timescale $\sim L/c$, particles with velocities predominantly aligned with $\hat{\boldsymbol{y}}$ take a longer time to cross the box and thus are accelerated to high energies. This broadens the global momentum distribution and forms a nonthermal tail.

\begin{figure}
\centerline{\includegraphics[width=\columnwidth]{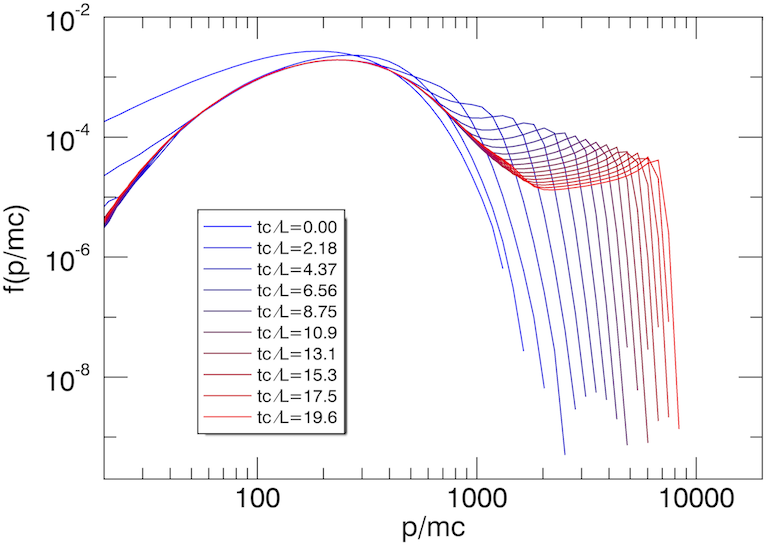}}
\centerline{\includegraphics[width=\columnwidth]{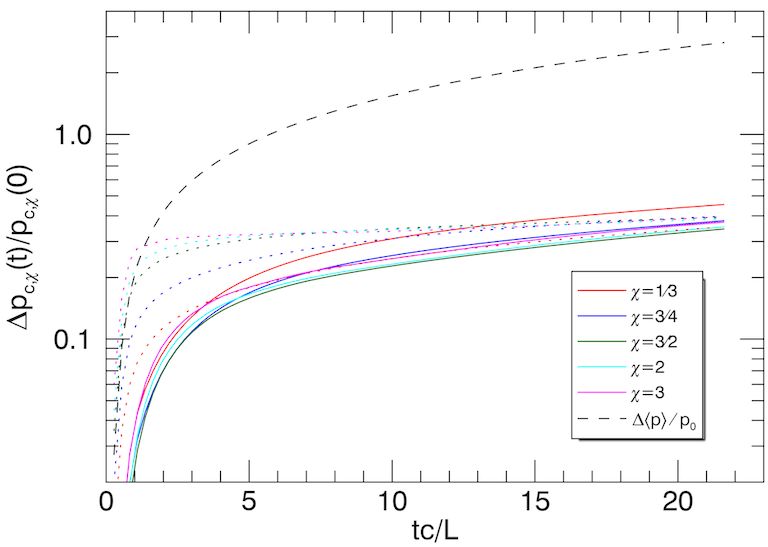}}
  \caption{Top panel: global 1D momentum distribution in the PIC simulation of a shear force acting on an uncharged collection of particles, demonstrating the formation of a nonthermal tail. Bottom panel: Change of the Casimir momenta $\Delta p_{{c},\chi}(t)$ relative to the initial value $p_{{c},\chi}(0)$ for the local coarse-grained distribution (solid lines) and for the global distribution (dotted lines); varying $\chi$ values are indicated by colors in the legend. For reference, the relative increase in the average momentum $\Delta \langle p \rangle(t)/\langle p \rangle(0)$ is shown in the black dashed line.}
\label{fig:evo_neutral}
\end{figure}

We show the evolution of the global 1D momentum distribution $\overline{f}(p,t)$ from the PIC simulation in the top panel of Fig.~\ref{fig:evo_neutral}; here, $\overline{f}(p,t) = p^2 \int d\Omega_p \overline{f}(\boldsymbol{p},t)$ where the integration is over directions of $\hat{\boldsymbol{p}}$. This momentum distribution is equivalent to the energy distribution since particles are relativistic ($E \approx p c$). The distribution forms a significant nonthermal tail with an extent that grows steadily with time, due to the (constantly depleted) subpopulation of particles that experience continuous acceleration.

The evolution of the relative change in the Casimir momenta, $\Delta p_{c,\chi}(t)/p_{c,\chi}(0) \equiv p_{c,\chi}(t)/p_{c,\chi}(0) - 1$, from this simulation is shown in the bottom panel of Fig.~\ref{fig:evo_neutral}, measured separately using both the local distribution $f(\boldsymbol{x},\boldsymbol{p},t)$ and the global distribution $\overline{f}(\boldsymbol{p},t)$; this is compared to the relative change in the average particle momentum $\Delta \langle p \rangle(t)/\langle p \rangle(0) \equiv \langle p \rangle(t)/\langle p \rangle(0) - 1$. The average particle momentum grows steadily by a factor of $\sim 3$ over the course of the simulation ($t \approx 22 L/c$) due to the constant shear acceleration. In contrast, $p_{c,\chi}$ only grow by $\sim 30\%$, indicating that while they are not conserved perfectly, they increase by a far smaller amount than $\langle p \rangle$. Thus, $p_{c,\chi}$ are {\it approximately conserved} in this case, despite the effect of phase mixing.

The evolution of $p_{c,\chi}$ is similar for all measured values of $\chi$ when using the local distribution, but exhibit differences for each $\chi$ when using the global distribution. In the latter case, $\Delta p_{c,\chi}(t)/p_{c,\chi}(0)$ for $\chi = 3$ grows rapidly before saturating at $\sim 0.3$ at the light crossing time $L/c$. For $\chi = 1/3$, the growth is much slower and only reaches the saturation value at the end of the simulation. Other values of $\chi$ are intermediate between the two extremes. This trend can be understood intuitively as follows. The dynamics are reversible on the transverse crossing timescale, $L/v_x$, for each particle, since the particles return to their initial state. Large values of $\chi$ (i.e., $\chi \gtrsim 1$) are sensitive to the thermal particles, which cross the box on the global light-crossing timescale, $\sim L/c$. Small values of $\chi$ (i.e., $\chi \lesssim 1$), on the other hand, are weighted toward particles in the tail of the distribution, which get accelerated to extreme energies because they have small $v_x$. These ``nonthermal'' particles take much longer to cross the box (i.e., their $L/v_x$ is large), so the dynamics do not repeat until late times. Thus, the saturation of $p_{c,\chi}$ reflects the timescale across which the dynamics are reversible for typical particles in the energy range that is captured by the weight $\chi$. In contrast, the evolution of the local distribution represents phase mixing on the local level, which always appears to be irreversible due to the existence of particles that have not yet crossed the box.

This uncharged system serves as a simple setting for understanding the entropy associated with free-streaming particles in a collisionless system. In this case, we conclude that {\it while entropy is produced {at coarse-grained scales} over time, it is far outpaced by the energization of the plasma}. This indicates that phase mixing by itself {enables} only a limited amount of entropy {production, and the outcome at late times is best represented by Scenario (3) in Sec.~\ref{sec:intro}}. We next increase the complexity by adding electromagnetic interactions.

\subsection{Parallel shear flow} \label{sec:parallel}

We next consider a magnetized plasma with a parallel shear force, $\boldsymbol{F}_{\rm shear} \parallel \boldsymbol{B}_0$. This situation is described by the full Vlasov-Maxwell equations {with force $\boldsymbol{F}_s = \boldsymbol{F}_{{\rm Lorentz},s} + \boldsymbol{F}_{\rm shear}$} including the Lorentz force $\boldsymbol{F}_{{\rm Lorentz},s} = q_s (\boldsymbol{E} + \boldsymbol{v} \times \boldsymbol{B}/c)$ (where $\boldsymbol{E}$ is the electric field, {$\boldsymbol{B}$ is the magnetic field, and $q_s$ is the electric charge for particle species $s$}) and the external shear force $F_{\rm shear}(x) \hat{\boldsymbol{y}}$ given by Eq.~\ref{eq:fysin}. {The guide field is thus $\boldsymbol{B}_0 = B_0 \hat{\boldsymbol{y}}$}.

Since particles stream freely along $\boldsymbol{B}_0$ while being unable to cross magnetic field lines beyond their gyroradius, to {lowest} order in $\rho_e/L$, the plasma at a given $x$ coordinate is simply advected along $\boldsymbol{B}_0$ at a speed that grows with time. {When electromagnetic field fluctuations are neglected and an isotropic initial distribution $f_0(p)$ is assumed, the analytical solution to the Vlasov equations for this setup is
\begin{align}
f(x,\boldsymbol{p},t) &= f_0\left( \sqrt{p_\perp^2 + (p_y - F_0 t \sin{(kx-\frac{k p_z c}{q_s B_0})})^2} \right) \nonumber \\
&\sim f_0\left( \sqrt{p_\perp^2 + (p_y - F_0 t \sin{(kx)})^2} \right) \label{eq:parsol}
\end{align}  
where the last expression holds in the limit of $k \rho_e \ll 1$, and we defined $p_\perp^2 = p_x^2 + p_z^2$.} In essence, a shear flow is continuously accelerated. This leads to a nonthermal population in the global distribution, since the plasma is locally accelerated to high energies in the regions of strong force. However, the local distribution remains a (shifted) thermal distribution. In other words, the global distribution is a superposition of drifting thermal distributions with a broad range of drift velocities.

\begin{figure}
\centerline{\includegraphics[width=\columnwidth]{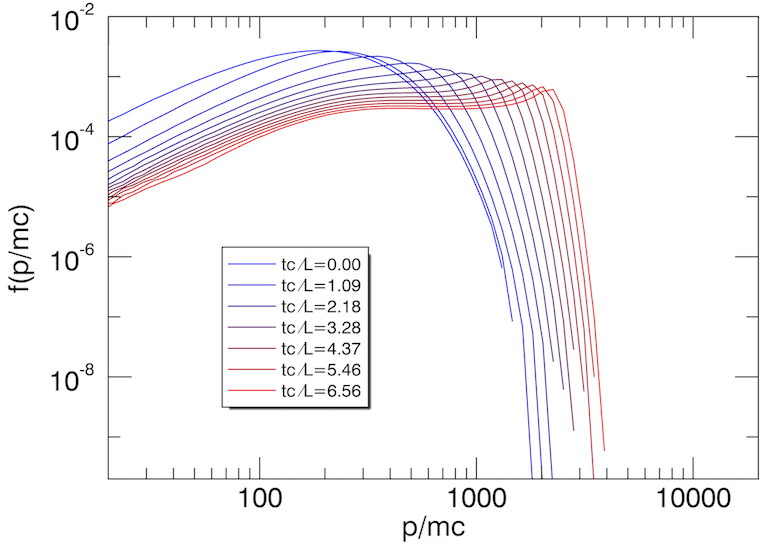}}
\centerline{\includegraphics[width=\columnwidth]{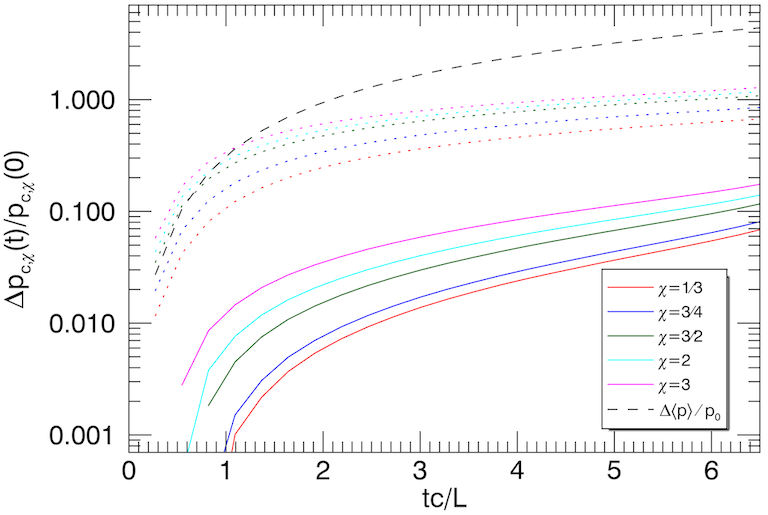}}
  \caption{Top panel: global 1D momentum distribution in the PIC simulation of a shear force acting parallel to the mean magnetic field $\boldsymbol{B}_0$, exhibiting broadening by bulk flows. Bottom panel: Change of the Casimir momenta $\Delta p_{{c},\chi}(t)$ relative to the initial value $p_{{c},\chi}(0)$ for the local coarse-grained distribution (solid lines) and for the global distribution (dotted lines); varying $\chi$ values are indicated by colors in the legend. For reference, the relative increase in the average momentum $\Delta\langle p \rangle(t)/\langle p \rangle(0)$ is shown in the black dashed line.}
\label{fig:evo_fyby}
\end{figure}

The evolution of the global 1D momentum distribution $\overline{f}(p,t)$ for the PIC simulation of the parallel shear flow is shown in the top panel of Fig.~\ref{fig:evo_fyby}. The shear flow acceleration produces a broad, flat distribution that grows to higher energies over time.

The evolution of the relative change in Casimir momenta, $\Delta p_{c,\chi}(t)/p_{c,\chi}(0)$, measured from the local and global distributions are shown in the bottom panel of Fig.~\ref{fig:evo_fyby}. For comparison, the mean momentum $\langle p \rangle$ increases by a factor of $\sim 5$ over the duration of $t \sim 6.5 L/c$; the injected energy is contained exclusively in the kinetic energy of the bulk flow, rather than internal energy. When $p_{{c},\chi}$ is measured from the local distribution, it increases by less than $20\%$ over this time scale. In fact, we find that the {conservation} of $p_{{c},\chi}$ is limited only by the numerical resolution: if we increase the resolution of the spatial bins while simultaneously increasing the number of particles to keep the same number of particles per bin, the growth of $p_{{c},\chi}$ can be further reduced (unlike the other cases described in this paper, which are converged{; see Appendix~\ref{sec:appendix2}}). In this sense, within the accuracy of our measurements, $p_{c,\chi}$ measured from the local coarse-grained distribution are consistent with being perfectly conserved in this case. Note that $p_{{c},\chi}$ grows by a smaller amount for $\chi = 1/3$ than for $\chi = 3$; this suggests that conservation is better represented by small values of $\chi$.

When $p_{{c},\chi}$ is measured from the global distribution $\overline{f}(\boldsymbol{p},t)$, we find that $p_{{c},\chi}$ increases by a factor of $\sim 2$ from the initial value, with an initial growth that is comparable to that of $\langle p \rangle$. Thus, there is an apparent entropy production in the global distribution due to the transfer of entropy from the spatial degrees of freedom (in this case, shear flow, which reduces the local entropy) to momentum-space degrees of freedom.

In summary, for a shear flow parallel to the background magnetic field, the Casimir momenta are {\it consistent with entropy conservation at the coarse-grained level, but exhibit significant entropy production at the global level}. {Thus, the outcome matches Scenario (2) in Sec.~\ref{sec:intro}.}

\subsection{Perpedicular shear flow} \label{sec:perp}

As the final laminar case, we consider a perpendicular shear force ($\boldsymbol{F}_{\rm shear} \perp \boldsymbol{B}_0$) by applying the external force $F_{\rm shear}(x) \hat{\boldsymbol{y}}$ given by Eq.~\ref{eq:fysin} and the guide field $\boldsymbol{B}_0 = B_0 \hat{\boldsymbol{z}}$. In this case, a shear flow is caused by $E \times B$ drift of particles across the magnetic field lines, rather than free streaming. As a result, an $\hat{\boldsymbol{x}}$ component of the electric field is established as the flow accelerates. An electric charge forms since $\nabla \cdot \boldsymbol{E} \neq 0$, the advection of which leads to an electric current $\boldsymbol{J}$ at wavenumber $2k$, and thus a weak fluctuation in the parallel magnetic field $\delta B_z$. While the early evolution of the flow is laminar, these effects {eventually} destabilize the flow {at times later than those considered here}.

\begin{figure}
\centerline{\includegraphics[width=\columnwidth]{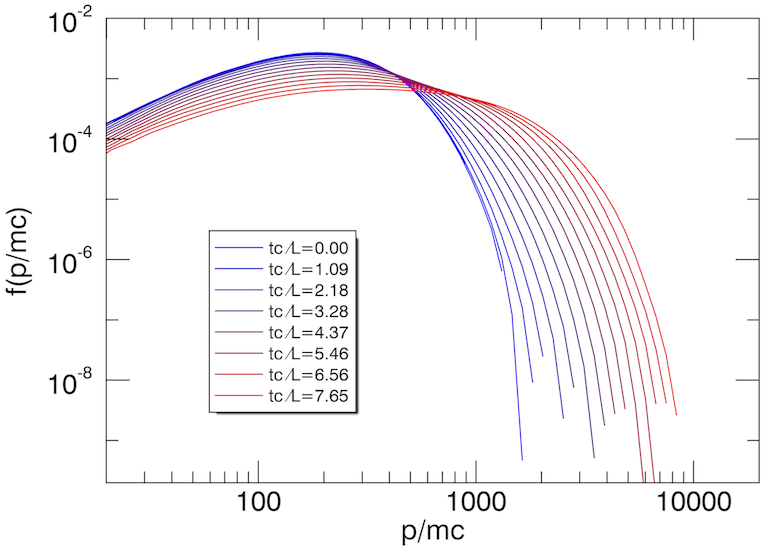}}
\centerline{\includegraphics[width=\columnwidth]{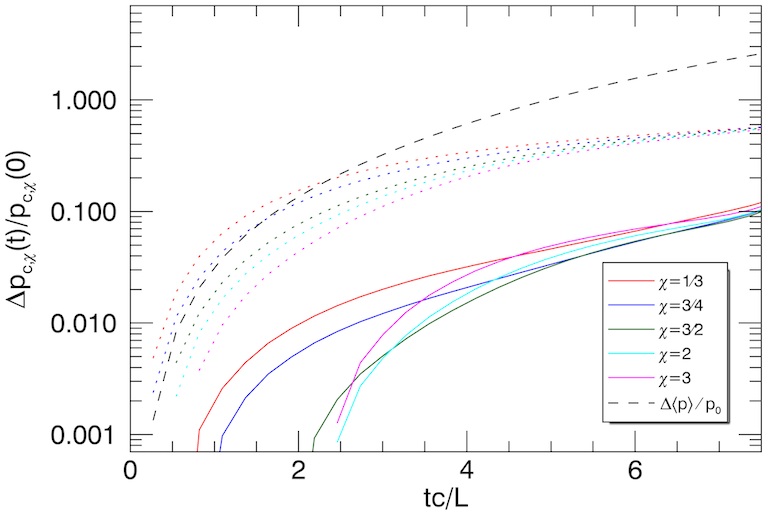}}
  \caption{Top panel: global 1D momentum distribution in the PIC simulation of a shear force acting perpendicular to the mean magnetic field $\boldsymbol{B}_0$, exhibiting broadening by bulk flows. Bottom panel: Change of the Casimir momenta $\Delta p_{{c},\chi}(t)$ relative to the initial value $p_{{c},\chi}(0)$ for the local coarse-grained distribution (solid lines) and for the global distribution (dotted lines); varying $\chi$ values are indicated by colors in the legend. For reference, the relative increase in the average momentum $\Delta\langle p \rangle(t)/\langle p \rangle(0)$ is shown in the black dashed line.}
\label{fig:evo_fybz}
\end{figure}

We show the evolution of the 1D momentum distribution from the PIC simulation of perpendicular shear flow in the top panel of Fig.~\ref{fig:evo_fybz}. Much like the previous cases, the distribution is broadened by the bulk flows, but this time it remains ``quasi-thermal'' in the sense of resembling a broadened thermal distribution.

The relative change in the Casimir momenta, $\Delta p_{c,\chi}(t)/p_{c,\chi}(0)$, is shown in the bottom panel of Fig.~\ref{fig:evo_fybz}, measured from both the local distribution and global distribution. Like the previous cases, $p_{c,\chi}$ from the local distribution grows with time, by $\sim 10\%$, which is a small change compared to $\langle p \rangle$, which increases by a factor of $\sim 4$ over the duration of the simulation ($t \sim 7.5 L/c$). The growth is similar for all values of $\chi$ considered, after a transient period during the first couple of light crossing times. One difference from the parallel shear flow is that the the local $p_{c,\chi}$ are converged with respect to binning in this case, indicating some degree of physical irreversibility. The evolution of $p_{c,\chi}$ from the global distribution is qualitatively similar as for the parallel shear flow, growing at a rate comparable to $\langle p \rangle$ initially before reaching an increase of $\sim 50\%$ for all $\chi$ at the end of the simulation. One difference from the previous case is that whereas $p_{c,\chi}$ grew the least for small values of $\chi$ in the parallel shear flow, it grows the least for larger values of $\chi$ in the perpendicular shear flow; this may be related to the different shapes of the distribution (nonthermal versus quasi-thermal). In this case, we conclude that {\it entropy is approximately conserved at the coarse-grained level, but produced in significant amounts at the global level}. {Thus, the outcome is again best represented by Scenario (2) in Sec.~\ref{sec:intro}.}

In summary, in all of the laminar shear flow cases, the growth of the Casimir momenta $p_{c,\chi}$ measured from the local ``coarse-grained'' distribution is on the order of $10\%$, which is about an order of magnitude less than the growth of the average momentum $\langle p \rangle$. This suggests that while entropy production is not entirely negligible, the process of bulk flow acceleration is predominantly entropy-conserving. When $p_{c,\chi}$ are measured with only the global distribution $\overline{f}$, there is a large amount of entropy production, comparable to $\langle p \rangle$, for all cases except the uncharged case. The non-conservation at the global level is not surprising since the Vlasov equations only predict the Casimir invariants of the local distribution to be conserved. With this intuition, we next consider the case of driven 2D turbulence.

\section{Numerical results on turbulent flow} \label{sec:turb}

\subsection{General evolution}

We now describe results from the turbulence simulation. We consider an out-of-plane mean field $\boldsymbol{B}_0 = B_0 \hat{\boldsymbol{z}}$. We drive turbulence via an external incompressible force that is perpendicular to $\boldsymbol{B}_0$, which consists of 20 modes with wavevectors $k_\perp \equiv (k_x^2 + k_y^2) \le 5^{1/2} 2\pi/L$ and independent randomly evolving phases. Essentially, a shear force is driven at each of these modes, similar to the one used to drive the laminar flows (Eq.~\ref{eq:fysin}) but with amplitude $F_0 = 2.4 T_0 k_\perp$ for each mode. The driving mechanism is described in more detail in our previous work~\citep{zhdankin_2021}.

\begin{figure}
  \centerline{\includegraphics[width=\columnwidth]{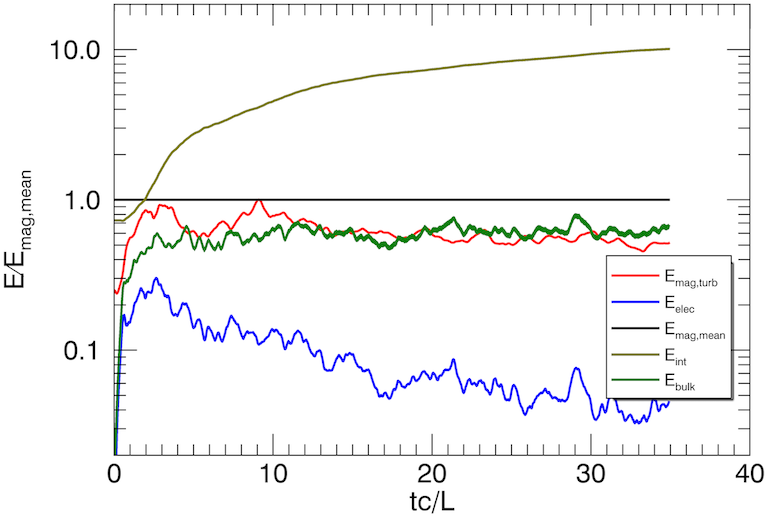}}
  \centerline{\includegraphics[width=\columnwidth]{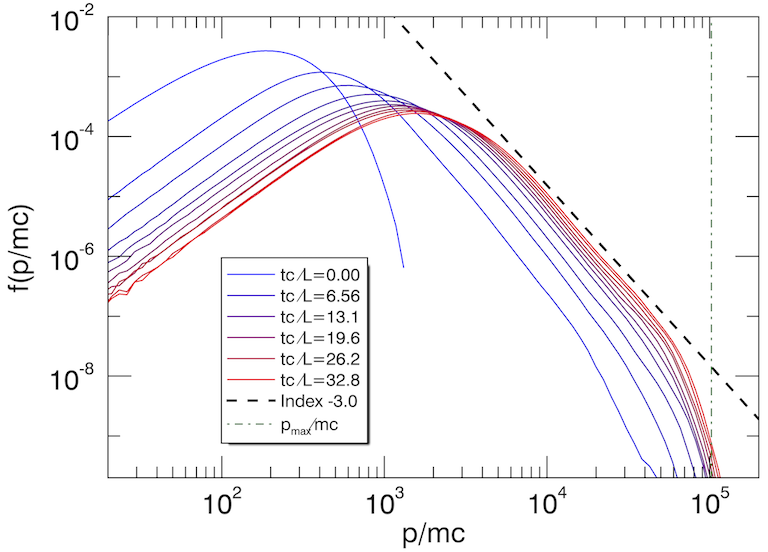}}
  \caption{Top panel: evolution of the global energies in the PIC simulation of 2D turbulence. Lines indicate turbulent magnetic energy (red), electric energy (blue), mean field magnetic energy (black), internal energy (gold), and turbulent bulk kinetic energy (green). Bottom panel: evolution of the global 1D particle momentum distribution, with a $p^{-3}$ power-law scaling shown for reference. The green dash-dotted line indicates the system-size limited momentum $p_{\rm max}$ at which the particle gyroradius is half of the box size.}
\label{fig:evo_turb}
\end{figure}

A spectrum of turbulent fluctuations quickly forms in the simulation. The evolution of the overall energy partition is shown in the top panel of Fig.~\ref{fig:evo_turb}. The fully developed turbulence is quasi-Alfv\'{e}nic (by construction), with the magnetic energy and bulk kinetic energies being comparable: $E_{\rm mag, turb} \sim E_{\rm bulk}$; the driving amplitude is such that both are roughly a factor of two less than the energy in the mean magnetic field $E_{\rm mag, mean}$. Note that there are no linear Alfv\'{e}n waves at large scales due to the absence of spatial variation along $\boldsymbol{B}_0$. The internal energy $E_{\rm int}$ (and thus $\langle p \rangle$) increases in time due to turbulent energy dissipation, growing by more than an order of magnitude over the course of the simulation (up to $tc/L \sim 35$). The electric energy $E_{\rm elec}$ is subdominant. We used the relativistic definitions for $E_{\rm bulk}$ and $E_{\rm int}$ described in our previous work~\citep{zhdankin_etal_2017}. Overall, the turbulence energies (and power spectra, not shown) are qualitatively similar to previously published 3D cases~\citep{zhdankin_etal_2017}, consistent with previous works~\citep[e.g.,][]{comisso_sironi_2018, comisso_sironi_2019}.

Relativistic turbulence leads to efficient nonthermal particle acceleration \citep[e.g.,][]{zhdankin_etal_2017, comisso_sironi_2018}. We show the evolution of the global 1D particle momentum distribution $\overline{f}(p,t)$ in the bottom panel of Fig.~\ref{fig:evo_turb}. Notably, a power-law tail with index $\alpha \approx -3$ is formed and maintained, in addition to some ``thermal'' heating that shifts the distribution peak to higher energies. Since an index of $\alpha = -3$ in the 1D momentum distribution corresponds to an index $a = 5$ in the 3D momentum distribution $f(\boldsymbol{p},t)$, this implies that the critical index for the Casimir momentum to be weighted toward the high-energy end of the tail is $\chi_{\rm crit} = 3/5$; values of $\chi$ smaller than this will capture primarily the nonthermal population (see Sec.~\ref{sec:pchar}).

\subsection{Evolution of Casimir momenta} \label{sec:turbpchar}

\begin{figure}
    \centerline{\includegraphics[width=\columnwidth]{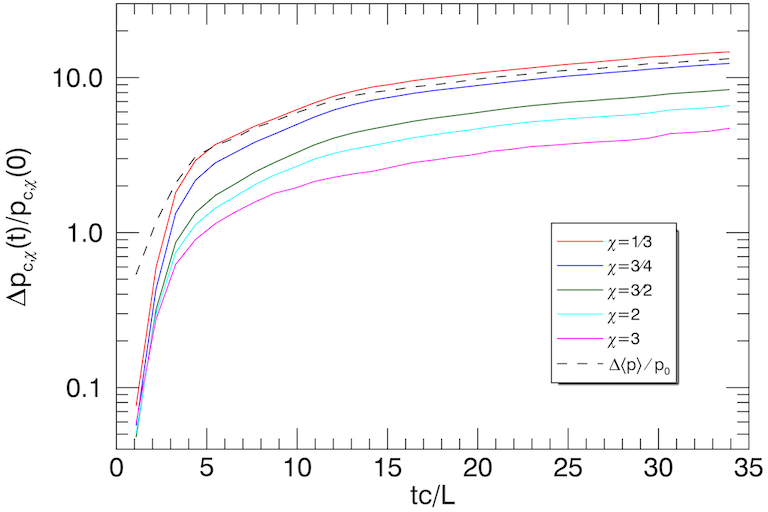}}
     \centerline{\includegraphics[width=\columnwidth]{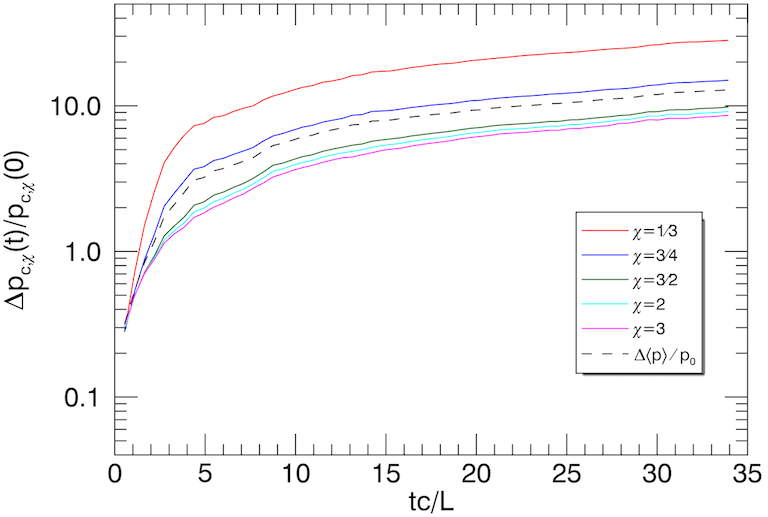}}
  \caption{Relative change of the Casimir momenta, $\Delta p_{{c},\chi}(t)/p_{{c},\chi}(0)$, computed using the local coarse-grained distribution (top panel) and the global distribution $\overline{f}$ (bottom panel); varying $\chi$ values are indicated by colors in the legend. For reference, the relative increase in the average momentum $\Delta\langle p \rangle(t)/\langle p \rangle(0)$ is shown in the black dashed line.}
\label{fig:entropy_turb}
\end{figure}

In the top panel of Fig.~\ref{fig:entropy_turb}, we show that in contrast to the laminar flows, the Casimir momenta $p_{{c},\chi}$ in the turbulence simulation grow at a rate comparable to that of the average momentum $\langle p \rangle$ for all $\chi$. In particular, we find that for $\chi < 1$, $\Delta p_{{c},\chi}(t)/p_{{c},\chi}(0) \approx \langle p \rangle(t)/\langle p \rangle(0) - 1$, indicating efficient thermalization. In this sense, turbulence is an efficient producer of entropy, despite the presence of the nonthermal tail in the global distribution. In contrast, for values of $\chi > 1$ weighted toward low energies, $p_{{c},\chi}$ remains a factor of few below $\langle p \rangle$, indicating better conservation at low energies than at high energies. Intuitively, the set of $p_{{c},\chi}$ evolve to cover a broad range of momenta, which indicates that the distribution evolves from a single characteristic energy (the thermal energy $\sim T$) to a distribution that spans over many energies.

The Casimir momenta measured from the global distribution $\overline{f}(\boldsymbol{p},t)$ (shown in the bottom panel of Fig.~\ref{fig:entropy_turb}) evolve in a similar way to those measured from the local (coarsed-grained) distribution, but the relative growth of $\Delta p_{{c},\chi}$ is larger by a factor of $\sim 2$. This is consistent with the considerations in Sec.~\ref{sec:casspatial} indicating that a uniform distribution will have higher $p_{c,\chi}$ (and entropy) than a spatially varying distribution. In this global case, $p_{{c},3/4}$ grows at a similar rate as $\langle p \rangle$, while $p_{{c},1/3}$ grows at an even faster rate than $\langle p \rangle$. This corresponds to $p_{c,\chi}$ inhabiting a range of values surrounding $\langle p \rangle$ within the power-law tail of the distribution. The value of $\chi = 1/3$ samples the high-energy end of the distribution (since $\chi_{\rm crit} = 3/5$ for $\alpha = -3$), consistent with the observation that $p_{{c},1/3}$ exceeds $\langle p \rangle$.

An important question is: do the growth rates of the Casimir momenta depend on the coarse-graining level relative to the kinetic scales and/or system size? To answer this, we compared the fiducial simulation to twice-smaller PIC simulations ($1024^2$ cells, $L/2\pi\rho_{e0} \approx 54$) with varying levels of coarse graining (and fixed number of macroparticles per spatial bin). When $\Delta x_{\rm bin}$ is twice smaller relative to the kinetic scales than the fiducial case, but the same size relative to the domain, the Casimir momenta from the smaller simulation evolve in a nearly identical manner to those from the fiducial simulation in Fig.~\ref{fig:entropy_turb}. Likewise, when $\Delta x_{\rm bin}$ is the same relative to the kinetic scales as the fiducial case, but twice smaller relative to the domain, the Casimir momenta also exhibit a nearly identical evolution. {We point the reader to Appendix~\ref{sec:appendix2} for details on this result.} We thus conclude that these results are converged with respect to coarse-graining level relative to either kinetic or global scales, for the range of $\Delta x_{\rm bin}$ considered.

In summary, {\it in the turbulence simulation, the evolution of $p_{c,\chi}$ is comparable to that of the average momentum $\langle p \rangle$, for both the local and the global distribution, indicating that a large amount of entropy is produced}. This result is in contrast to the laminar flows, demonstrating the link between irreversible energization (via the turbulent cascade) and entropy production. {The situation is then consistent with Scenario (1) in Sec.~\ref{sec:intro}, although one cannot {\it a priori} rule out the possibility of Scenario (2) holding at finer scales; we will return to this point in Sec.~\ref{sec:conclusions}.}

\subsection{Spatial profile of local Casimir momenta}

We close this work by commenting on the spatial structure of the Casimir momenta measured in local subdomains of the turbulence simulation, as a preliminary consideration of their potential as a diagnostic for local irreversible energy dissipation. By measuring the Casimir invariants in each spatial bin, rather than integrating over the entire domain, we can construct local Casimir momenta, denoted $p^{\rm loc}_{{c},\chi}(\boldsymbol{x},t)${, defined in the same way as Eq.~\ref{eq:pchar} but with the spatial integral constrained to bins of size $\Delta x_{\rm bin}$ (while keeping the normalization factors based on the global average number density $n_0$ and using $N \to n_0 V_{\rm bin}$ where $V_{\rm bin}$ is the bin volume). In the continuum limit ($\Delta x_{\rm bin} \to 0$), this would become
\begin{align}
p^{\rm loc}_{{c},\chi}(f) &\to n_0^{1/3} \left( \frac{1}{n_0} \int d^3p f^\chi  \right)^{-1/3(\chi-1)} \, . \label{eq:pcharloc}
\end{align}
}Since each subdomain has open boundaries, the local Casimir momenta $p^{\rm loc}_{{c},\chi}$ are not conserved by the Vlasov equation {due to advection effects}, but may give insight into structures which contribute the most to the overall Casimir momenta $p_{{c},\chi}$.

\begin{figure*}
\includegraphics[width=\columnwidth]{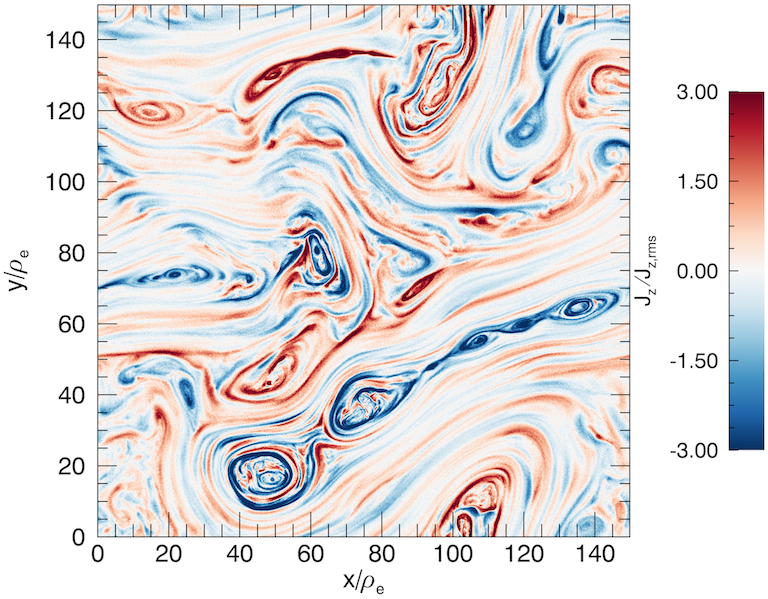}
\includegraphics[width=\columnwidth]{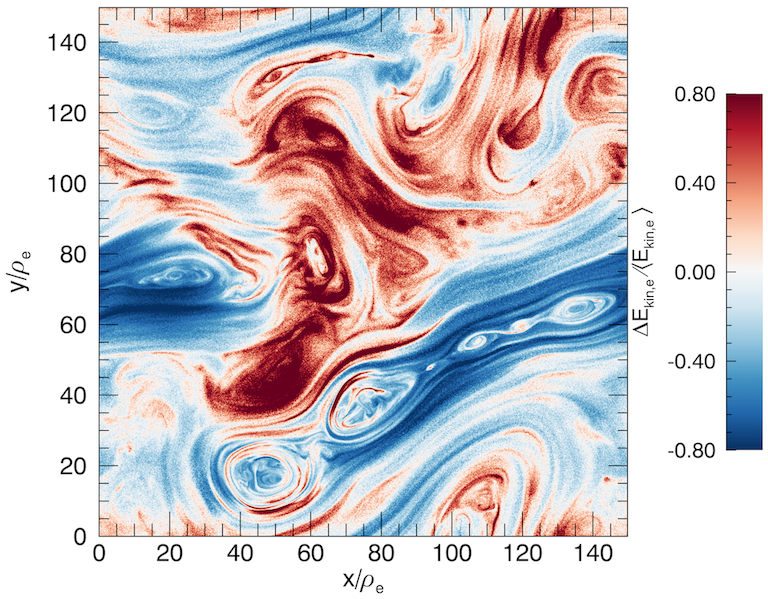}
\includegraphics[width=\columnwidth]{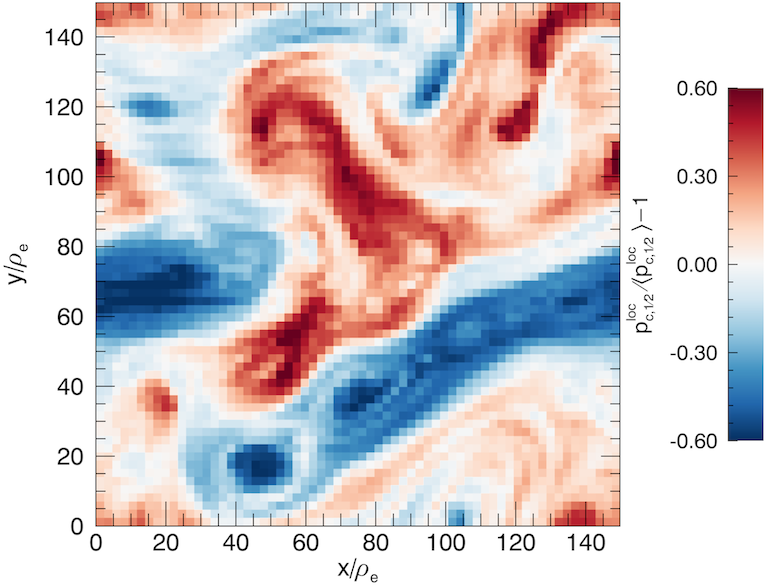}
\includegraphics[width=\columnwidth]{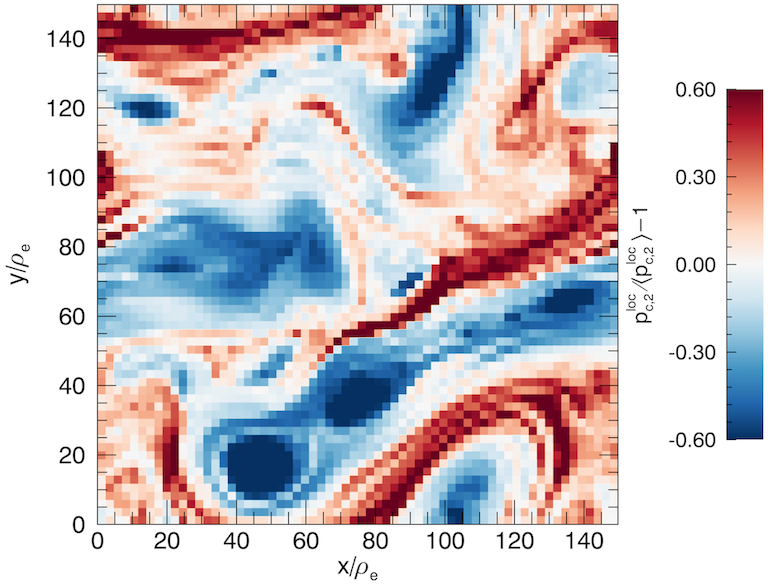}
    \caption{{Clockwise from top left panel: the spatial profile of the out-of-plane current density $J_z$ (relative to the rms value), fluctuations in the electron energy density $E_{{\rm kin},e}$ (relative to the mean value), and local Casimir momenta $p^{\rm loc}_{c,\chi}$ (normalized to the mean value) for $\chi = 2$ and $\chi = 1/2$ in the turbulence simulation.}}
\label{fig:images_turb}
\end{figure*}

{In the top panels of Fig.~\ref{fig:images_turb}}, we show images of the out-of-plane electric current density $J_z$ and electron kinetic energy density {$E_{{\rm kin},e}$}, at an arbitrary time ($tc/L = 8.75$) during fully developed turbulence. Intermittent current sheets and plasmoids formed by magnetic reconnection are evident in the current density; much of the high-energy plasma, however, resides outside of such structures. {In the bottom panels of Fig.~\ref{fig:images_turb}}, we show the spatial profile of $p^{\rm loc}_{{c},\chi}$ for $\chi = 1/2$ and $\chi = 2$ (acquired from the electron distribution).  We find that $p^{\rm loc}_{{c},1/2}$ is very strongly correlated with {$E_{{\rm kin},e}$}, confirming that the Casimir momenta at low $\chi$ are associated with regions of heated plasma. Curiously, while $p^{\rm loc}_{{c},2}$ also exhibits some correlation with {$E_{{\rm kin},e}$}, the dominant structures have a different morphology and location than $p^{\rm loc}_{{c},1/2}$. In particular, the largest values of of $p^{\rm loc}_{{c},2}$ occur in filamentary structures upstream of the reconnection layers and plasmoids. The qualitative differences in the spatial profile of $p^{\rm loc}_{{c},1/2}$ and $p^{\rm loc}_{{c},2}$ confirm that the different values of $\chi$ are sensitive to different dissipation processes.

Recent work has suggested that in relativistic turbulence, low-energy particles may be preferentially energized by magnetic reconnection in intermittent current sheets, while high-energy ones may be energized by diffusive particle acceleration \citep[see discussion in Ref.][]{comisso_sironi_2019}. One may anticipate that some of the differences between $p^{\rm loc}_{{c},1/2}$ and $p^{\rm loc}_{{c},2}$ {are} related to this process: since $p^{\rm loc}_{{c},2}$ is sensitive to lower-energy particles than $p^{\rm loc}_{{c},1/2}$, it will be more sensitive to irreversible dissipation processes occurring at those low energies.

This concludes our application of the Casimir momenta for measuring entropy production in PIC simulations of 2D relativistic turbulence. We next summarize our results and point out these future directions.

\section{Conclusions} \label{sec:conclusions}

In this work, we created a theoretical framework for understanding the anomalous production of kinetic entropy in collisionless plasmas by manipulating the more general Casimir invariants. We introduced the infinite set of Casimir momenta to characterize the violation of entropy conservation (in a generalized sense) relative to the energy injection; these momenta are defined simply by
\begin{align}
p_{{c},\chi} \equiv n_0^{1/3} \left( \frac{1}{N} \int d^3x d^3p f^\chi  \right)^{-1/3(\chi-1)} \, , \label{eq:conc}
\end{align}
where $\chi$ is a free index that parameterizes the weight toward different regions of phase space (low energy for large $\chi$ and high energy for small $\chi$). The Vlasov equation predicts $p_{{c},\chi}$ to be conserved, for suitable boundary conditions and in the absence of collisions, but anomalous entropy production ({enabled by} phase mixing or nonlinear entropy cascades) breaks the conservation of $p_{{c},\chi}$.

This theoretical framework provides a novel approach to characterizing entropy in collisionless plasmas, bypassing ambiguities with the standard {BG} entropy (in particular, the {non-uniqueness, arbitrary normalization, and arbitrary zero point}). In the author's opinion, further investigation of the mathematical properties of the Casimir momenta and their evolution in various analytical problems is warranted. {The Casimir invariants have played a key role in previous analytical works by constraining the properties of possible solutions to the Vlasov-Maxwell equations \cite[such as in the proof of the existence of weak solutions by Ref.][]{diperna_lions_1989}; the Casimir momenta may provide a means to better characterize departures of the coarse-grained dynamics from the Vlasov equation.}

As an initial application, we employed this theoretical framework to PIC simulations of 2D relativistic kinetic turbulence, to demonstrate that a substantial amount of entropy is produced despite nonthermal features in the particle distribution. {This rules out Scenario (3) described in Sec.~\ref{sec:intro}.} At a glance, this appears to limit the direct application of the Casimir invariants to constrain the form of the nonthermal particle distribution after turbulent particle energization. {However, it opens up the possibility of using maximum-entropy principles to explain certain aspects of nonthermal distributions, which will be explored in future work. We also} foresee that measurements of the Casimir momenta may lead to a better understanding of the kinetic processes responsible for irreversible energy dissipation, by rigorously characterizing the competition between entropy production and entropy conservation. In particular, the Casimir momenta measured in local subdomains may serve as a proxy for identifying the spatial structures responsible for irreversible dissipation. In future work, it is desirable to perform a more comprehensive numerical analysis by studying the spatial structure and statistics of the Casimir momenta in more detail and in 3D geometry. 

An important theoretical question is whether the anomalous entropy production observed in our turbulence simulation is due to phase mixing or due to a nonlinear entropy cascade. {{Broadly speaking,} the former {allows an increase of} entropy at coarse scales by accumulating {structure} at finer scales, as described by Scenario (2) in Sec.~\ref{sec:intro}, while the latter {forms singularities that enable} collisional entropy production via Scenario (1).} The neutral shear flow (described in Sec.~\ref{sec:neutral}) reveals that entropy production from pure phase mixing {of counterstreaming flows} has only a limited contribution to entropy production. {The phase mixing of density fluctuations may also contribute to entropy production (as described in Appendix~\ref{sec:appendix3}); however, this channel is likely subdominant for incompressible driving (and $\boldsymbol{B}_0$ oriented out of the 2D domain).} {Given these considerations}, it is reasonable to surmise that {a significant contribution to} the anomalous entropy production {in turbulence} comes from the entropy cascade{, responsible for the thermal heating of the distribution at low energies}. {The fact that growth rates of the Casimir momenta are nearly identical when considering simulations with varying coarse-graining scales {(as shown in Appendix~\ref{sec:appendix2}) and system sizes {may be viewed as further support for this scenario}. {However, we again emphasize that Scenario (1) and Scenario (2) may be difficult (if not impossible) to distinguish if both processes persist to fine sub-kinetic scales.} In future work, {it may be useful to consider PIC simulations with explicit collisions in order to control and diagnose directly the contribution of Scenario (1).} It will also need to be demonstrated whether {our findings extend} to other parameter regimes (such as low or high $\beta$) and to 3D geometry, where parallel phase mixing might be enhanced due to spatial variability along the global guide field.}

The Casimir momenta derived in this paper may be useful for characterizing local heating in weakly collisional plasmas such as the solar wind and Earth's magnetosphere, where {high-quality} information of the particle momentum distribution is available from {\it in situ} spacecraft measurements. Indeed, recent work has compared various energy-based and distribution-based measures of irreversible dissipation, and highlighted the need for more rigorous methods \citep{pezzi_etal_2021}. The Casimir momenta provide one such distribution-based measure, which may have some conceptual advantages over other measures such as the {BG} entropy.

The Casimir momenta may be applied to better understand dissipative plasma processes other than turbulence. An obvious candidate for this is magnetic reconnection. Recent studies have applied numerical simulations to study the reversibility of magnetic reconnection, and suggested that it may be reversible for sufficiently strong guide field \citep[e.g.,][]{ishizawa_watanabe_2013, xuan_etal_2021}. Likewise, there is numerical evidence for the {BG} entropy being conserved to high precision in PIC simulations \citep{liang_etal_2019}. If so, the Casimir momenta may provide a useful constraint on reduced modeling of the evolved plasma distribution.

The theoretical framework presented in this paper (based on dimensional representations of generalized entropy) may be relevant for a range of statistical mechanical problems outside of plasma physics. One example is in galactic dynamics, where stellar distribution functions are modeled via the gravitational Vlasov-Poisson equation. Previous theoretical works have suggested using Casimir invariants as a basis for generalized statistics \citep[see e.g.,][]{lynden-bell_1962, lynden-bell_1967, tremaine_etal_1986}, whereas numerical studies of $N$-body systems have mainly focused on characterizing anomalous BG entropy production rates and mechanisms \citep[e.g.,][]{esilva_etal_2017, esilva_etal_2019, esilva_etal_2019b}; Casimir momenta (or related dimensional invariants) may provide a method for bridging the two. {Recent analytical work has also proposed extensions of generalized statistics to plasmas \citep{ewart_etal_arxiv} and other collisionless systems with long-range interactions (e.g., dark matter) \citep{chavanis_arxiv}, which have yet to be tested by numerical simulations.} Future work {will} assess the applicability of our ideas to these various problems. {The possibility of using the Casimir momenta as a basis for generalized statistics is further studied in a follow-up paper to this work \citep{zhdankin_2022}, which applies maximum-entropy principles to derive power-law nonthermal distributions from the Casimir momenta.}

{The author would like to thank the {three} anonymous referees of this manuscript for multiple suggestions that improved the manuscript.} The author acknowledges support for this work from NASA through the NASA Hubble Fellowship grant \#HST-HF2-51426.001-A awarded by the Space Telescope Science Institute, which is operated by the Association of Universities for Research in Astronomy, Inc., for NASA, under contract NAS5-26555. Research at the Flatiron Institute is supported by the Simons Foundation. This work used the Extreme Science and Engineering Discovery Environment (XSEDE), which is supported by National Science Foundation grant number ACI-1548562. This work used the XSEDE supercomputer Stampede2 at the Texas Advanced Computer Center (TACC) through allocation TG-PHY160032 \citep{xsede}.

\appendix

\section{Example of a uniform entropy-conserving distribution} \label{sec:appendix}

It may be helpful for the reader to consider an example of a distribution that systematically changes its shape while conserving all of the Casimir invariants. For this purpose, consider an initially uniform plasma with an isotropic distribution $f_0(p)$. Suppose that energy is injected into the plasma, in a way such that the plasma ultimately remains uniform (perhaps after a transient period) but an anisotropy and nonthermal population are produced (the details of the force $\boldsymbol{F}$ required to accomplish this will not be considered). Assume that the anisotropy is cylindrically symmetric about $p_z$ (which may be, e.g., the direction of the background magnetic field). In this case, an entropy-conserving solution is given by
\begin{align}
f(\boldsymbol{x},\boldsymbol{p},t) = f_0\left( \sqrt{a p_\perp^2 + p_z^2/a^2} \right) \, ,
\end{align}
where $p_\perp^2 = p_x^2 + p_y^2$ and $a(t)$ is a rescaling factor such that $a(0) = 1$, determined by the amount of energy injected (i.e., details of $\boldsymbol{F}$ and duration). The Casimir invariants (Eq.~\ref{eq:casg}) are conserved since
\begin{align}
{\mathfrak C}_g(f) &= \frac{2\pi}{N} \int d^3x dp_\perp dp_z p_\perp g\left(f_0(\sqrt{a p_\perp^2 + p_z^2/a^2})\right) \nonumber \\
&= \frac{2\pi}{N} \int d^3x dp'_\perp dp'_z p'_\perp g\left(f_0\left( \sqrt{{p'_\perp}^2 + {p'_z}^2} \right)\right) \nonumber \\
&= {\mathfrak C}_g(f_0) \, ,
\end{align}
where we defined $p'_\perp = a^{1/2} p_\perp$ and $p'_z = p_z/a$. In this solution, the plasma effectively cools in the perpendicular direction and heats in the parallel direction if $a > 1$, and vice versa if $a < 1$. The resulting anisotropy scales with the amount of injected energy. If $f_0$ is a relativistic Maxwell-J\"{u}ttner, i.e., $f_0 \propto \exp{(-p/p_0)}$ where $p_0$ is the initial thermal momentum, the resulting energy distribution will be highly nonthermal (i.e., a hard power law, with an extent determined by $a$). Although this example is very idealized, it demonstrates that the Casimir invariants (and thus entropy) can be conserved {at macroscopic scales} by simple distributions, in principle. 

{\section{Dependence on bin size} \label{sec:appendix2}}

{In this section, we present some results from our convergence study of the Casimir momenta $p_{c,\chi}$ with respect to distribution bin size in the spatial dimension, $\Delta x_{\rm bin}$. We performed PIC simulations that are identical to the neutral shear flow from Sec.~\ref{sec:neutral}, parallel shear flow from Sec.~\ref{sec:parallel}, perpendicular shear flow from Sec.~\ref{sec:perp}, and turbulent flow from Sec.~\ref{sec:turb}, except that $\Delta x_{\rm bin}$ is varied from the fiducial value ($\Delta x_{\rm bin} = L/32$ for the laminar flows and $\Delta x_{\rm bin} = L/64$ for the turbulent flow) while the number of macroparticles per bin is held fixed.  All simulations are well converged with respect to number of particles per cell. In Fig.~\ref{fig:convergence}, we show the relative change in the Casimir momenta, $\Delta p_{c,\chi}(t)/p_{c,\chi}(0)$, for $\chi \in \{ 1/3, 3 \}$ and varying coarse-graining scale $\Delta x_{\rm bin}$. We also show corresponding quantity computed from the global distribution (marked $\Delta x_{\rm bin} = L$) with the fidicual number of macroparticles per cell.}

{For the neutral shear flow (top left panel), we take $\Delta x_{\rm bin} \in \{ L/32, L/16, L/8 \}$; the cases with $\Delta x_{\rm bin} \le L/16$ show no significant dependence on $\Delta x_{\rm bin}$. For the parallel shear flow (top right panel), we take $\Delta x_{\rm bin} \in \{ L/64, L/32, L/16 \}$; this case does not exhibit any signs of convergence with $\Delta x_{\rm bin}$, which is consistent with $p_{c,\chi}$ being perfectly conserved in the fine-grained limit. For the perpendicular shear flow (bottom left panel), we take $\Delta x_{\rm bin} \in \{ L/64, L/32, L/16 \}$, and find that results are essentially converged when $\Delta x_{\rm bin} \le L/32$. Finally, for the turbulence case, we take $\Delta x_{\rm bin} \in \{ L/64, L/32, L/16 \}$, and find that results are converged when $\Delta x_{\rm bin} \le L/32$. The results indicate that the asymptotic/converged $\Delta p_{c,\chi}(t)/p_{c,\chi}(0)$ is relatively small in the laminar cases, but large in the turbulence case. One notable observation is that for the neutral shear flow and turbulence cases, $\Delta p_{c,\chi}(t)/p_{c,\chi}(0)$ appears to converge faster for $\chi = 1/3$ compared to $\chi = 3$.}

\begin{figure*}
\includegraphics[width=\columnwidth]{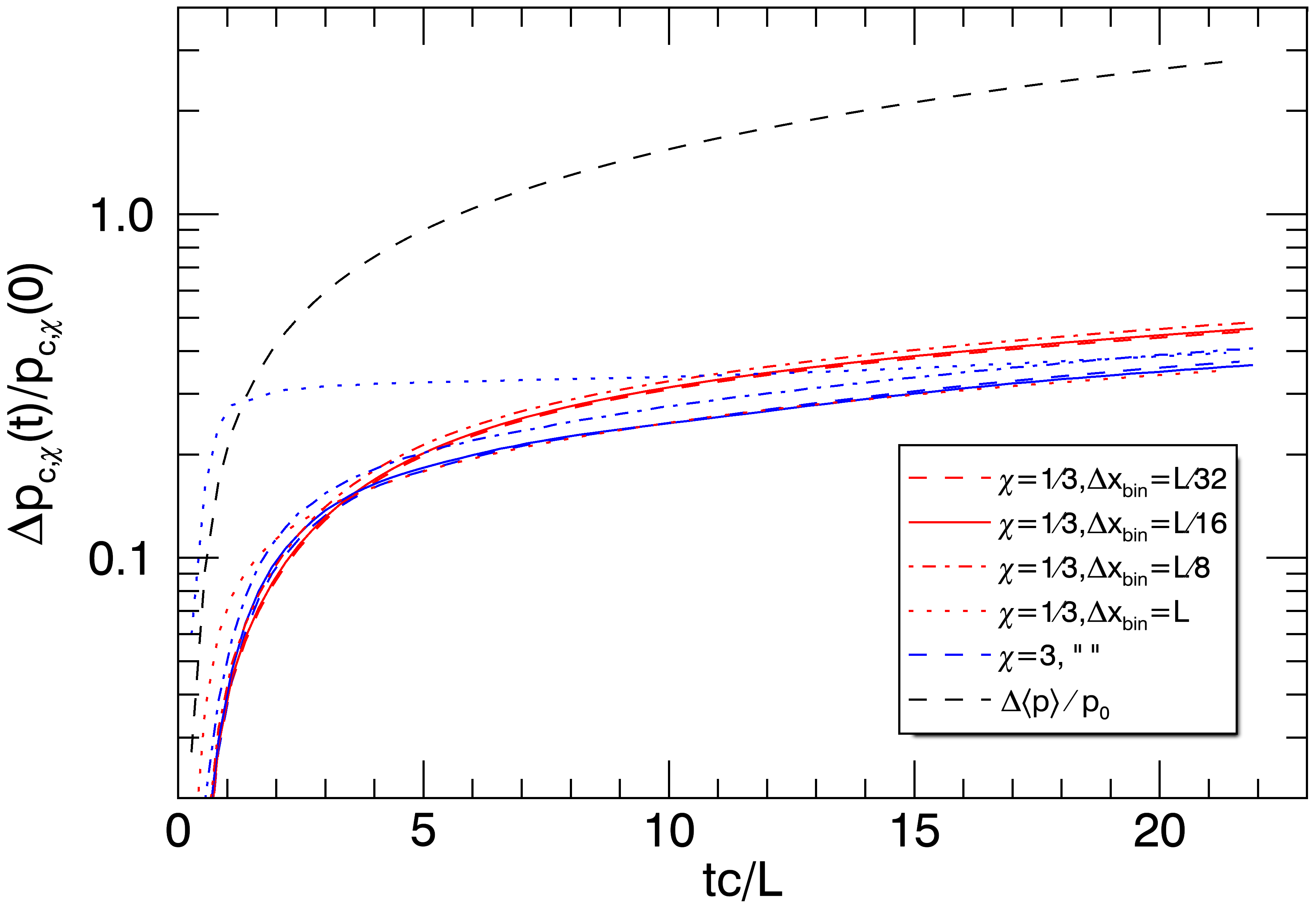}
\includegraphics[width=\columnwidth]{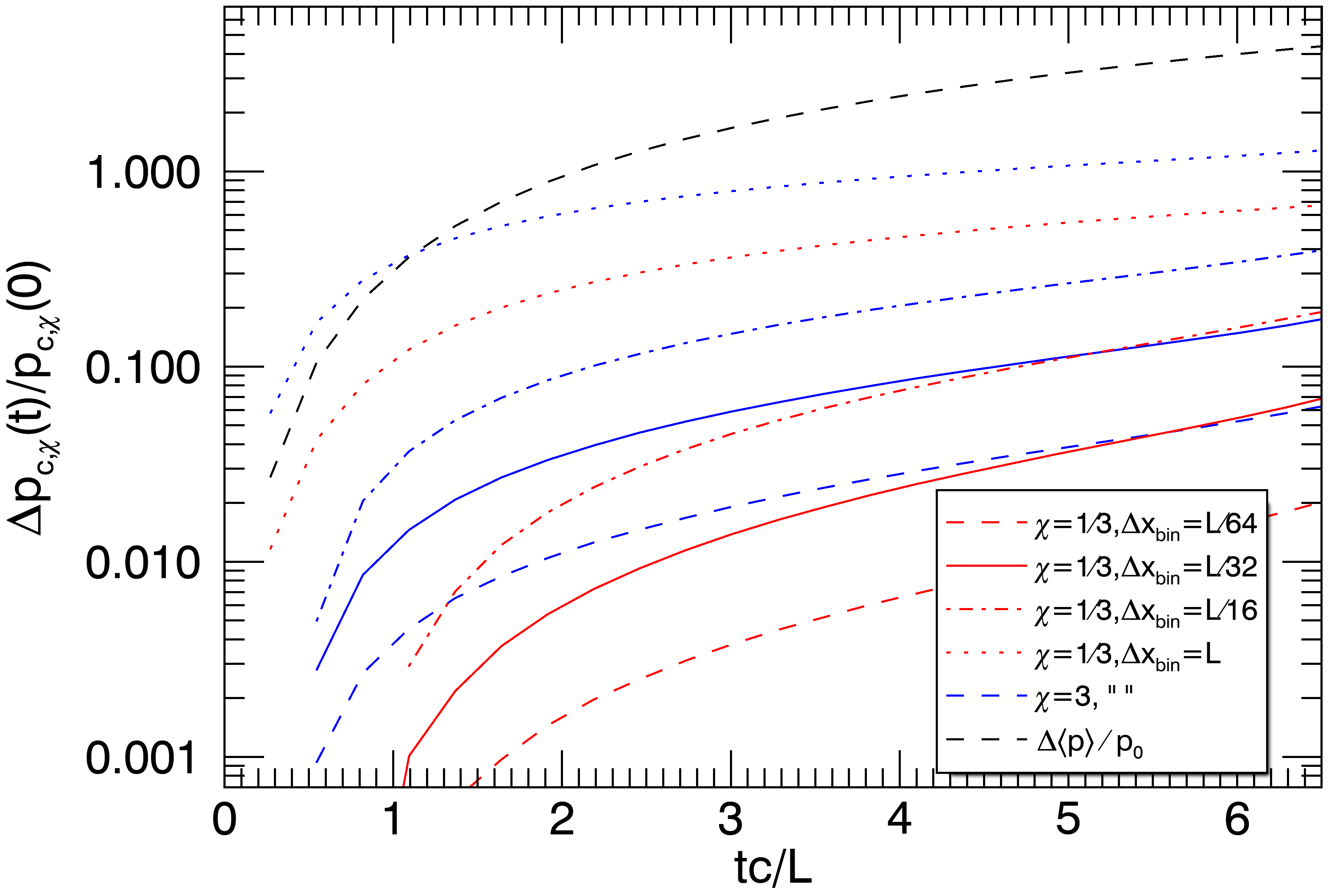}
\includegraphics[width=\columnwidth]{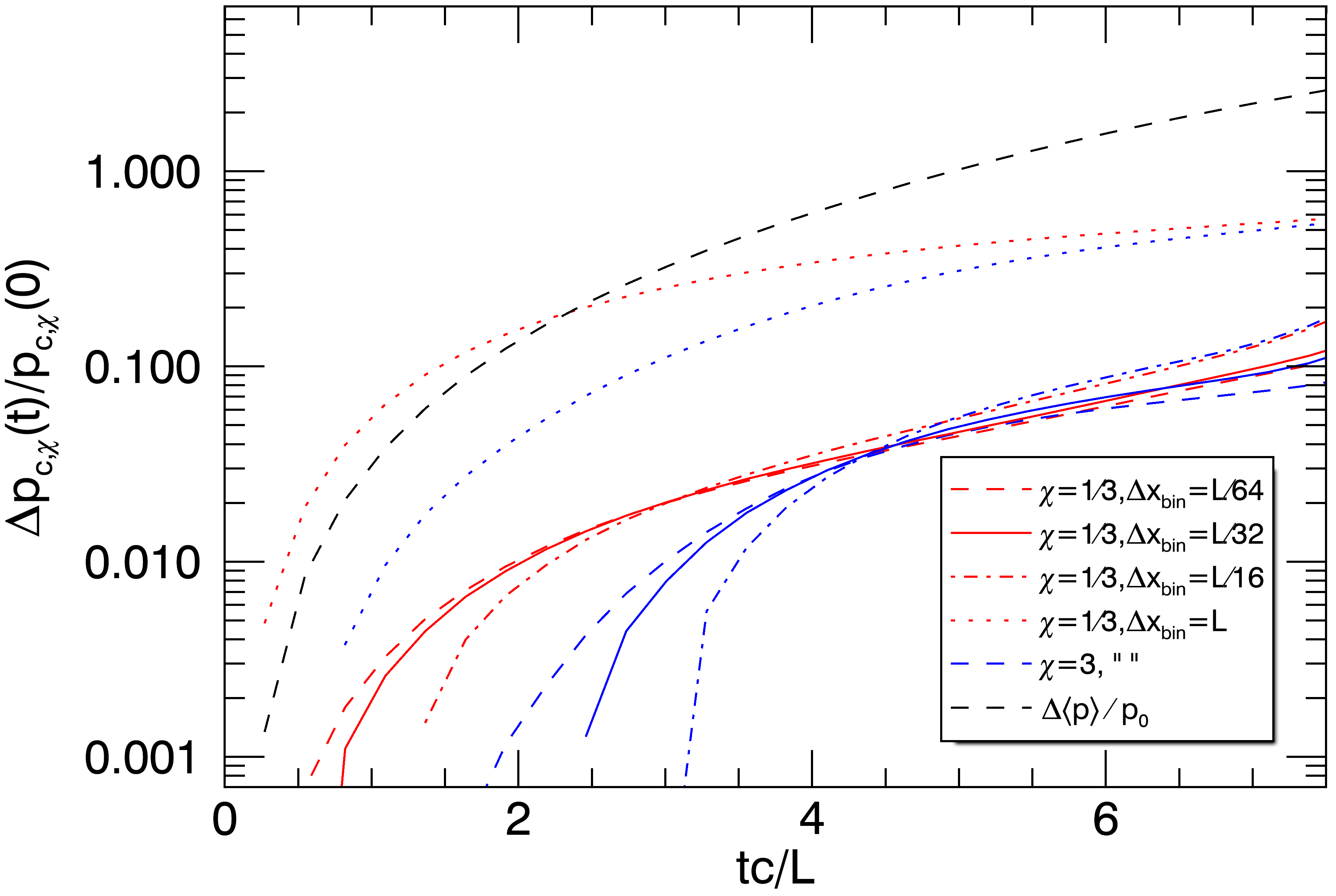}
\includegraphics[width=\columnwidth]{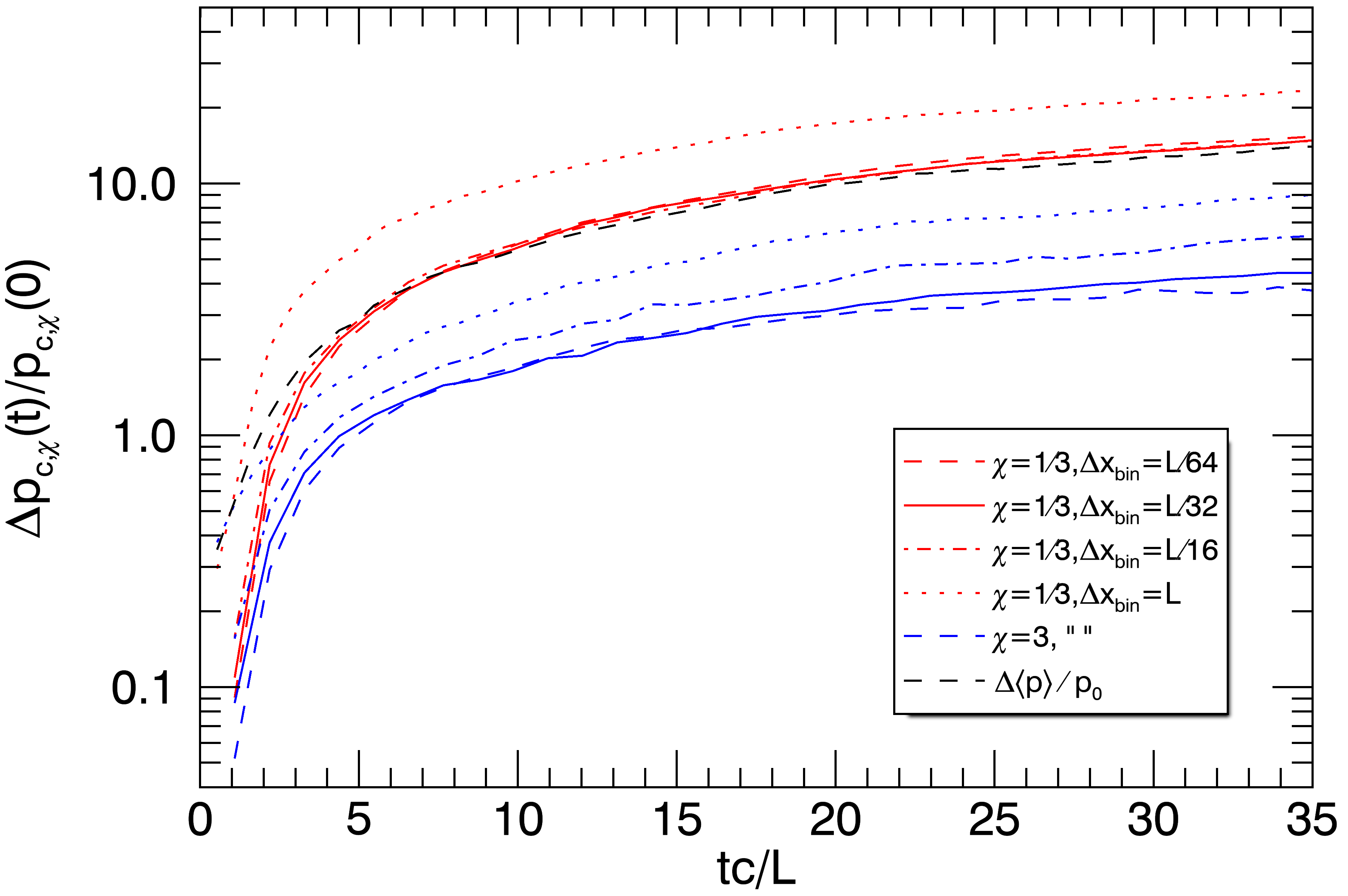}
  \caption{{Change of the Casimir momenta $\Delta p_{{c},\chi}(t)$ relative to the initial value $p_{{c},\chi}(0)$ for $\chi = 1/3$ (red) and $\chi = 3$ (blue), with varying coarse-graining levels (indicated by linestyles in the legend, including the global distribution indicated by $\Delta x_{\rm bin} = L$). The panels correspond to the neutral shear flow from Sec.~\ref{sec:neutral} (top left), parallel shear flow from Sec.~\ref{sec:parallel} (top right), perpendicular shear flow from Sec.~\ref{sec:perp} (bottom left), and turbulent flow from Sec.~\ref{sec:turb} (bottom right). For comparison, the relative change in the average momentum $\Delta \langle p \rangle / p_0$ is also shown (black dashed line).}}
\label{fig:convergence}
\end{figure*}

{Finally, we show that the results are insensitive to the system size for the turbulence case. In Fig.~\ref{fig:convergence2}, we compare $\Delta p_{c,\chi}(t)/p_{c,\chi}(0)$ (with $\chi \in \{ 1/3, 3 \}$) for the fiducial ($L/2\pi\rho_e \approx 109$, $\Delta x_{\rm bin} = L/64$) turbulence simulation with two twice-smaller ($L/2\pi\rho_e \approx 54$) cases that have $\Delta x_{\rm bin} = L/64$ and $\Delta x_{\rm bin} = L/32$. The similarity of these three cases supports the statements made in Sec.~\ref{sec:turbpchar}.}

\begin{figure}
\includegraphics[width=\columnwidth]{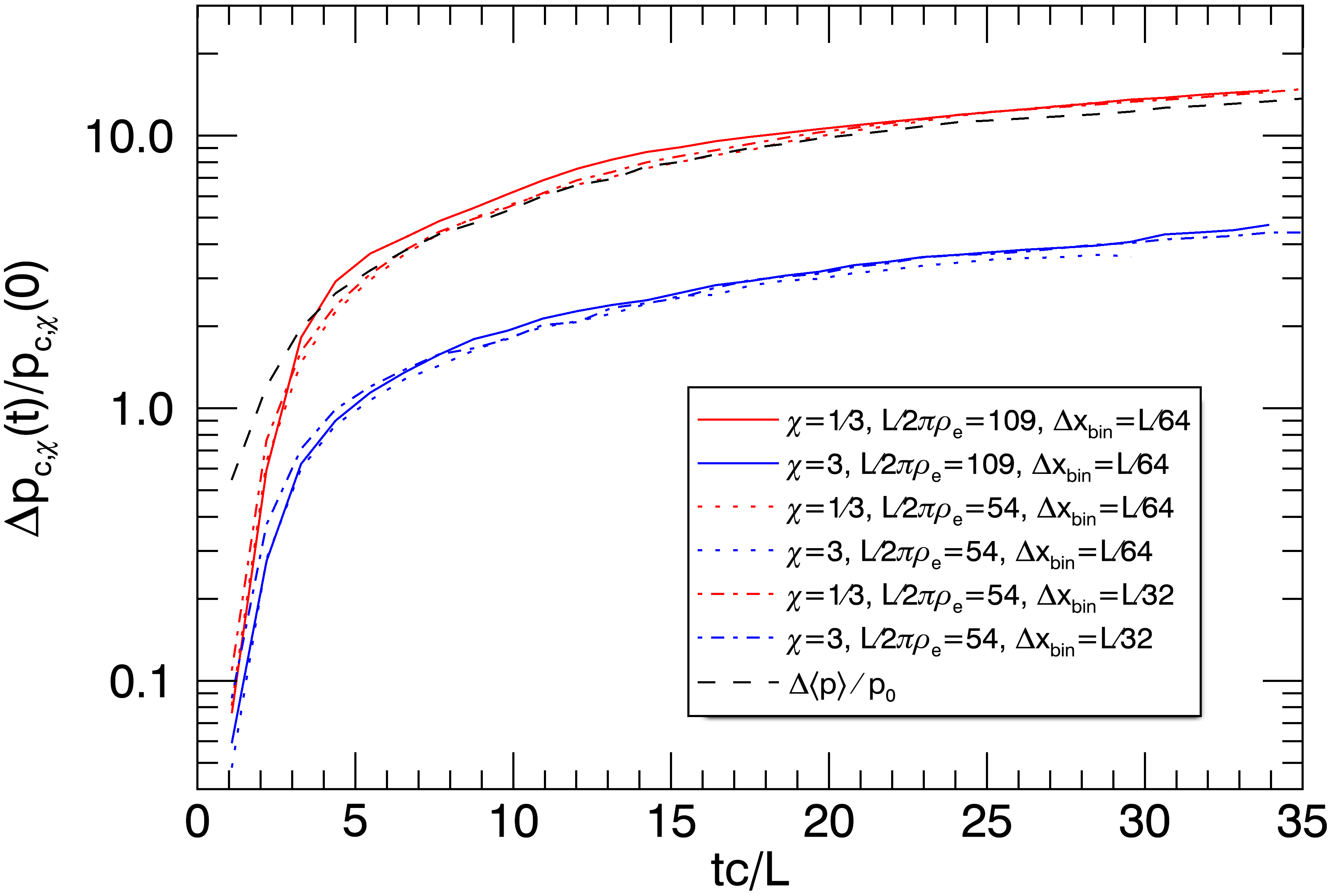}
  \caption{{Change of the Casimir momenta $\Delta p_{{c},\chi}(t)$ relative to the initial value $p_{{c},\chi}(0)$ for $\chi = 1/3$ (red) and $\chi = 3$ (blue), for the fiducial ($L/2\pi\rho_e \approx 109$, $\Delta x_{\rm bin} = L/64$) turbulence simulation (solid line), similar case with $L/2\pi\rho_e \approx 54, \Delta x_{\rm bin} = L/64$ (dotted line), and similar case with $L/2\pi\rho_e \approx 54, \Delta x_{\rm bin} = L/32$ (dash-dotted line). For comparison, the relative change in the average momentum $\Delta \langle p \rangle / p_0$ is also shown (black dashed line).}}
\label{fig:convergence2}
\end{figure}

{\section{Damping of density perturbation} \label{sec:appendix3}}

{The collisionless damping of a density perturbation is one of the simplest examples of coarse-grained entropy production via phase mixing. In this situation, there is no energy injected into the particles, so the global particle distribution is preserved during the decay and the average particle momentum is fixed to the initial value, $\langle p \rangle(t) = p_0$. Nevertheless, the Casimir momenta measured from the coarse-grained distribution experience growth due to the development of fine structure in the local distribution. In this section, we demonstrate that this growth is small compared to that observed in the turbulence simulation of Sec.~\ref{sec:turb}.}

{Consider a pair plasma immersed in a uniform magnetic field $\boldsymbol{B}_0=B_0 \hat{\boldsymbol{x}}$, with a density perturbation $\delta n(x)$ that has spatial variation only along $\boldsymbol{B}_0$. We assume an initial distribution (for each species) of the form $f_0(x,\boldsymbol{p}) = [1 + \delta n_0(x)/n_0] f_g(p)$ where $n_0$ is the mean number density and $f_g(p)$ is the global momentum distribution, taken to be isotropic. The Vlasov equation for either species reduces to an advection equation $\partial_t f + v_x \partial_x f = 0$ (exploiting gyrosymmetry and species mass symmetry), which has a solution
\begin{align}
f(x,\boldsymbol{p},t) &= f_0(x - v_x t, \boldsymbol{p}) \nonumber \\
&= [1 + \delta n_0(x-v_x t)/n_0] f_g(p) \label{eq:fdens} \, .
\end{align}
A similar solution would be obtained for a density fluctuation in a neutral collisionless gas. Eq.~\ref{eq:fdens} indicates that the distribution at each value of $v_x$ is simply advected by that velocity. Over long times, the initial fluctuation is sheared apart in phase space; upon coarse graining, the distribution approaches the uniform $f_g(p)$ at asymptotically late times. If we assume perfect mixing, then we can relate the Casimir momenta of the initial state to those of the final state,
\begin{align}
p_{c,\chi,0} &= n_0^{1/3} \left[ \frac{1}{N} \int d^3x d^3p f_0^\chi \right]^{-1/3 (\chi-1)} \nonumber \\
&= n_0^{1/3} \left[ \frac{1}{N} \int d^3x \left( 1 + \frac{\delta n_0}{n_0} \right)^\chi \int d^3p f_g^\chi \right]^{-1/3 (\chi-1)} \nonumber \\
&= \left[ \langle \left(1 + \frac{\delta n_0}{n_0}\right)^{\chi} \rangle\right]^{-1/3(\chi-1)} p_{c,\chi,\infty} \, , \label{eq:pchidens}
\end{align}
where brackets indicate a spatial average and $p_{c,\chi,\infty}$ are the asymptotic Casimir momenta computed from $f_g$. Thus, the phase-mixed final state is of higher entropy, $p_{c,\chi,\infty} > p_{c,\chi,0}$, as long as the distribution is coarse-grained at any level below the initial fluctuation scale (regardless of the presence or absence of collisions).}

{As a concrete example, consider a periodic domain that is initially half empty ($n = 0$) and half full ($n = 2 n_0$), with the profile $n(x) = 2 n_0 H(x - L/2)$ where $H$ is the Heaviside step function. In this case, $\langle (1 + \delta n_0/n_0)^\chi \rangle = 2^{\chi-1}$ so that Eq.~\ref{eq:pchidens} implies $p_{c,\chi,\infty} = 2^{1/3} p_{c,\chi,0}$ for all $\chi$. Thus, the decay of the density fluctuation produces only a modest increase of $p_{c,\chi}$, despite the high amplitude. To confirm that the asymptotic value is reached on a reasonable timescale, we performed a PIC simulation with this density profile and similar parameters to the other laminar flow simulations in this work (relativistic pair plasma in 2D domain, but no external driving and a larger $\Delta p_{\rm bin}$ by factor of 4). The resulting evolution of $p_{c,\chi}$ for several values of $\chi$ is shown in Fig.~\ref{fig:densitydecay}, confirming that $\Delta p_{c,\chi}/p_{c,\chi,0} \to 2^{1/3} - 1 \approx 0.26$ (to a fair numerical accuracy) on the timescale of a few light crossing times.}

\begin{figure}
\centerline{\includegraphics[width=\columnwidth]{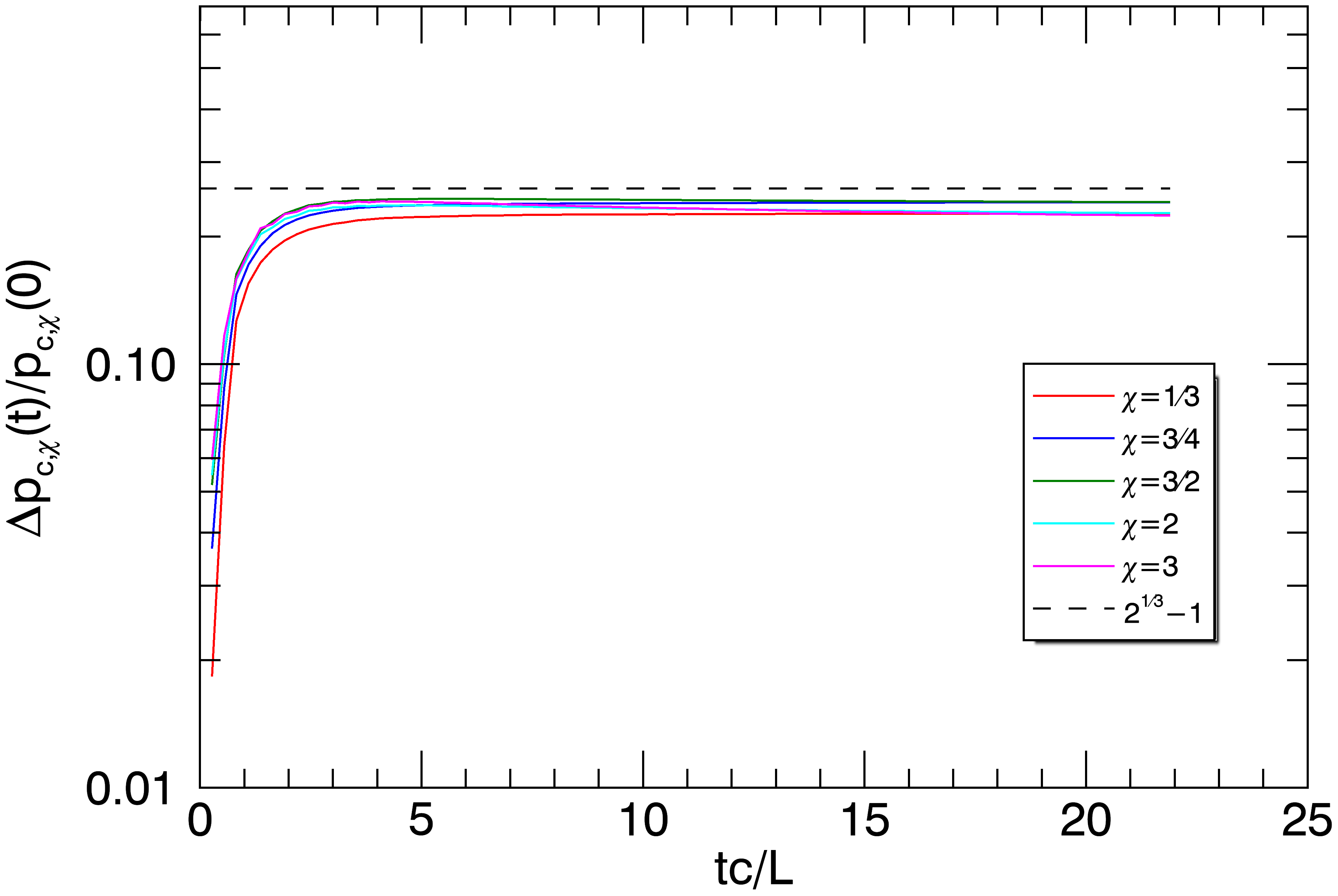}}
  \caption{{Change of the Casimir momenta $\Delta p_{{c},\chi}(t)$ relative to the initial value $p_{{c},\chi}(0)$ (from the local coarse-grained distribution) measured for the decay of a density fluctuation; varying $\chi$ values are indicated by colors in the legend. For reference, the asymptotic limit of $\Delta p_{c,\chi}(t)/p_{c,\chi}(0) \to 2^{1/3} - 1$ from phase mixing is indicated by the dashed line.}}
\label{fig:densitydecay}
\end{figure}


%

\end{document}